\documentclass[iop]{emulateapj}

\usepackage{braket}

\shorttitle{Planet Traps and the Planet-Metallicity Relation}
\shortauthors{Hasegawa \& Pudritz}

\begin{document}

\title{Planet Traps and Planetary Cores: Origins of the Planet-Metallicity Correlation}

\author{Yasuhiro Hasegawa\altaffilmark{1}}
\affil{Institute of Astronomy and Astrophysics, Academia Sinica (ASIAA), P.O. Box 23-141, Taipei 10641, Taiwan}
\email{yasu@asiaa.sinica.edu.tw}

\author{Ralph E. Pudritz\altaffilmark{2}}
\affil{Department of Physics and Astronomy, McMaster University,
    Hamilton, ON L8S 4M1, Canada}
\email{pudritz@physics.mcmaster.ca}

\altaffiltext{1}{EACOA fellow}

\altaffiltext{2}{Origins Institute, McMaster University, Hamilton, ON L8S 4M1, Canada}

\begin{abstract}
Massive exoplanets are observed preferentially around high metallicity ([Fe/H]) stars while low-mass exoplanets do not show such an effect.  
This so-called planet-metallicity correlation generally favors the idea
that most observed gas giants at $r<10$ AU are formed via a core accretion process.
We investigate the origin of this phenomenon using a semi-analystical model, 
wherein the standard core accretion takes place at planet traps in protostellar disks where rapid type I migrators are halted.
We focus on the three major exoplanetary populations 
- hot-Jupiters, exo-Jupiters located at $r \simeq 1$ AU, and the low-mass planets.  
We show using a statistical approach that the planet-metallicity correlations are well reproduced in these models. 
We find that there are specific transition metallicities with values [Fe/H]$=-0.2$ to $-0.4$, 
below which the low-mass population dominates, and above which the Jovian populations take over. 
The exo-Jupiters significantly exceed the hot-Jupiter population at all observed metallicities.  
The low-mass planets formed via the core accretion are insensitive to metallicity, 
which may account for a large fraction of the observed super-Earths and hot-Neptunes.
Finally, a controlling factor in building massive planets is the critical mass of planetary cores ($M_{c,crit}$) 
that regulates the onset of runaway gas accretion.  
Assuming the current data is roughly complete at [Fe/H]$>-0.6$, 
our models predict that the most likely value of the "mean" critical core mass 
of Jovian planets is $\braket{M_{c,crit}}  \simeq 5 M_{\oplus}$ rather than $10 M_{\oplus}$.  
This implies that grain opacities in accreting envelopes should play an important role in lowering $M_{c,crit}$.

\end{abstract}


\section{Introduction} \label{intro}

\begin{figure*}
\begin{minipage}{17cm}
\begin{center}
\includegraphics[width=8cm]{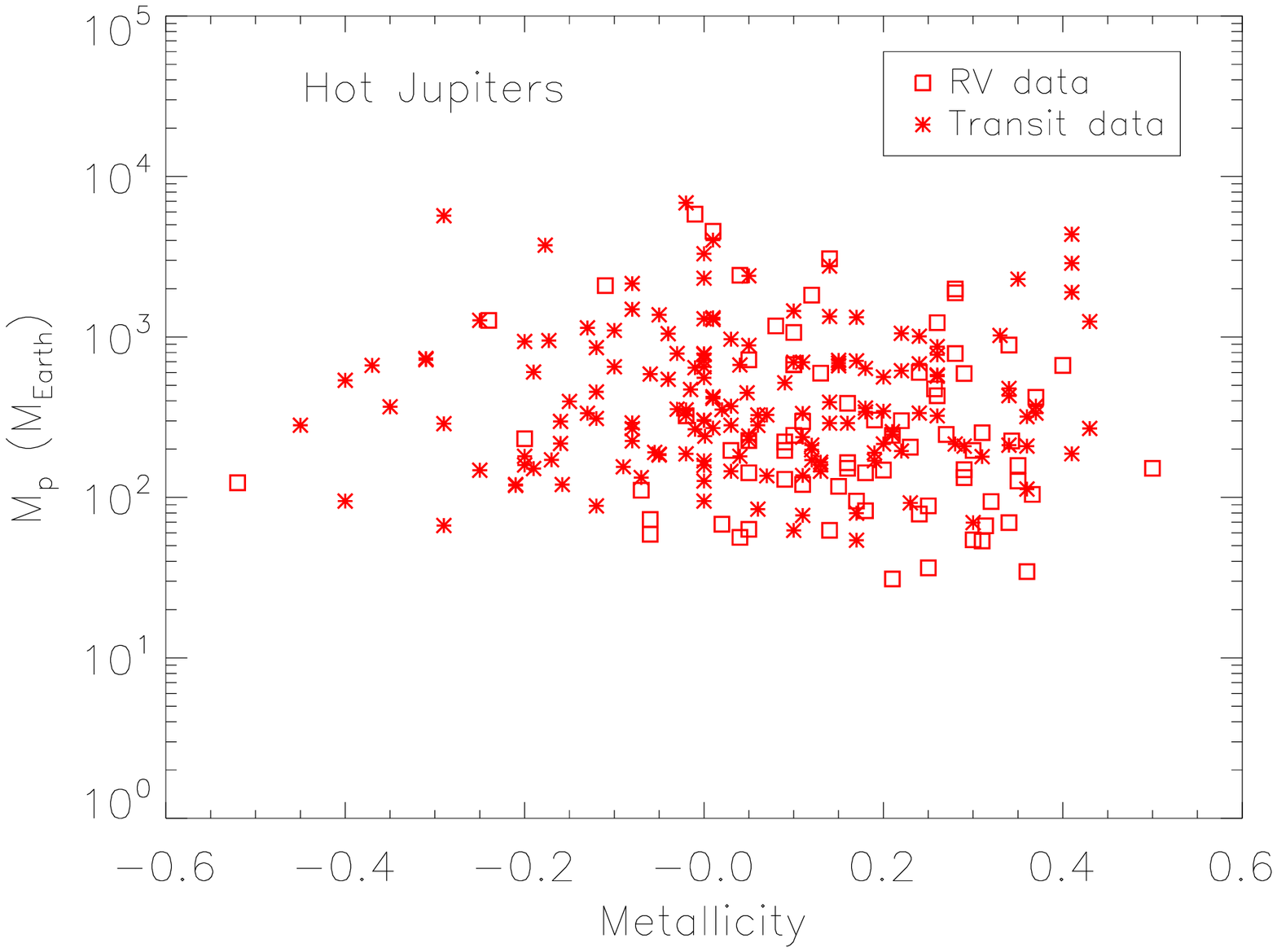}
\includegraphics[width=8cm]{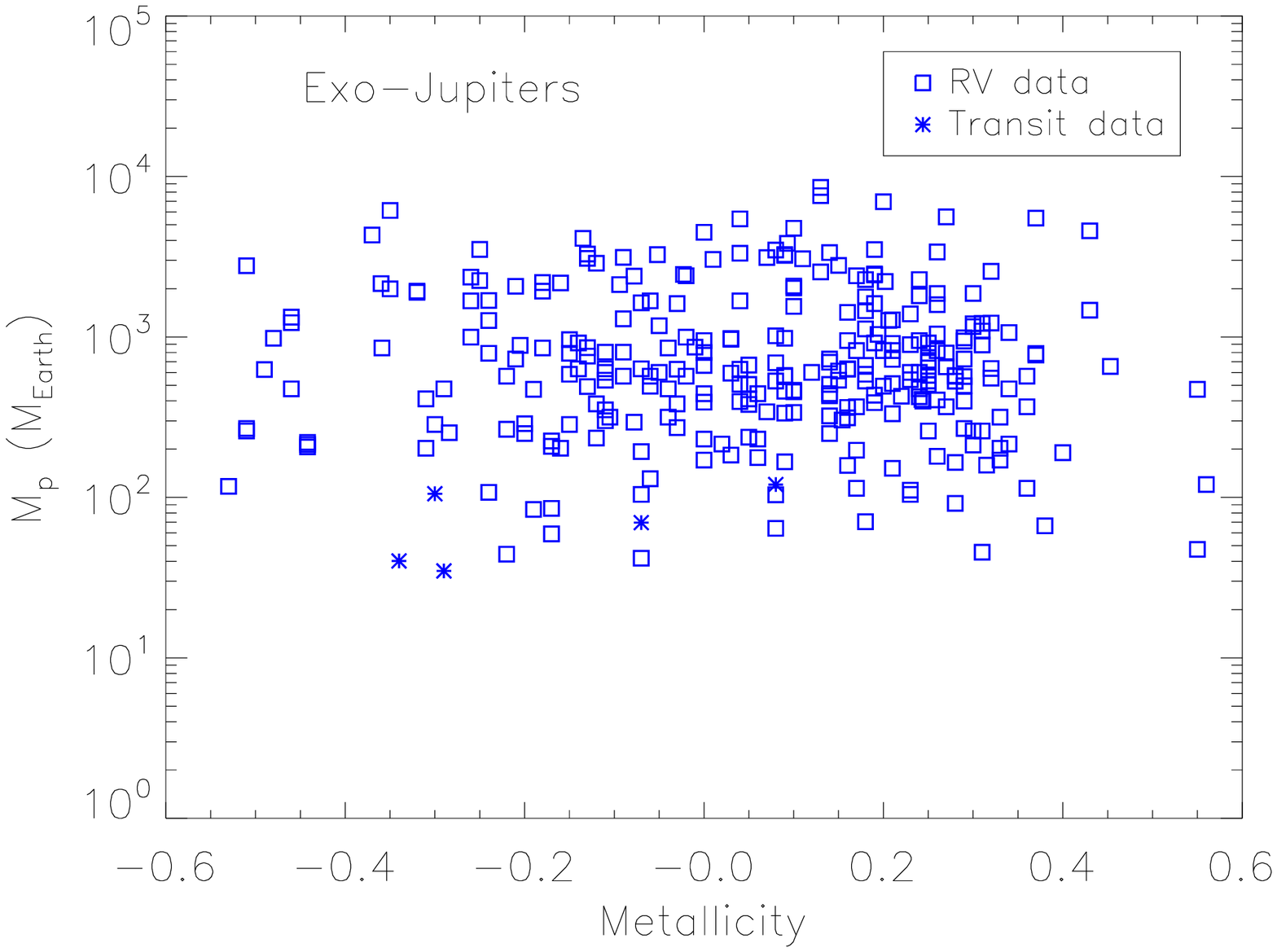}
\includegraphics[width=8cm]{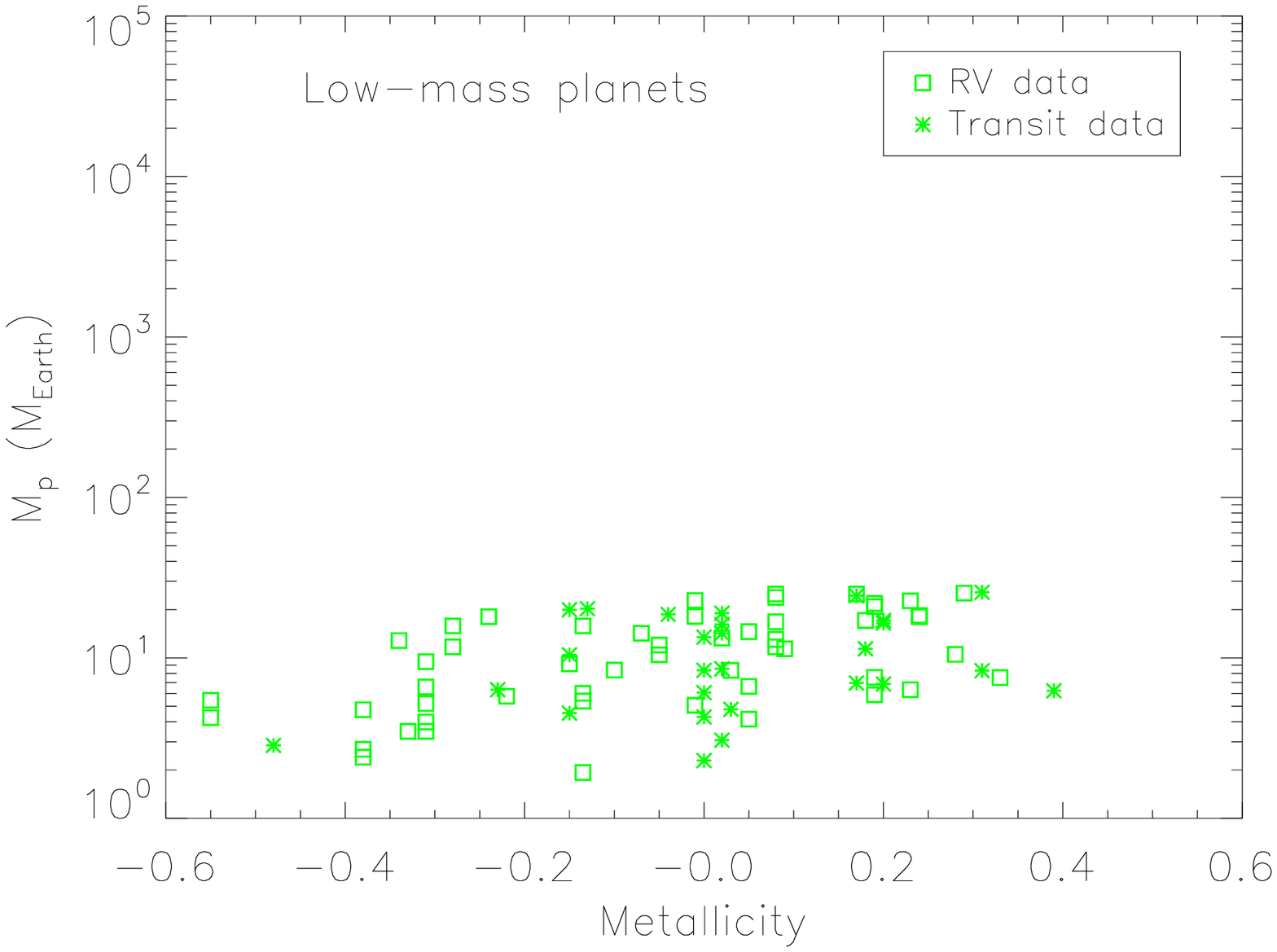}
\includegraphics[width=8cm]{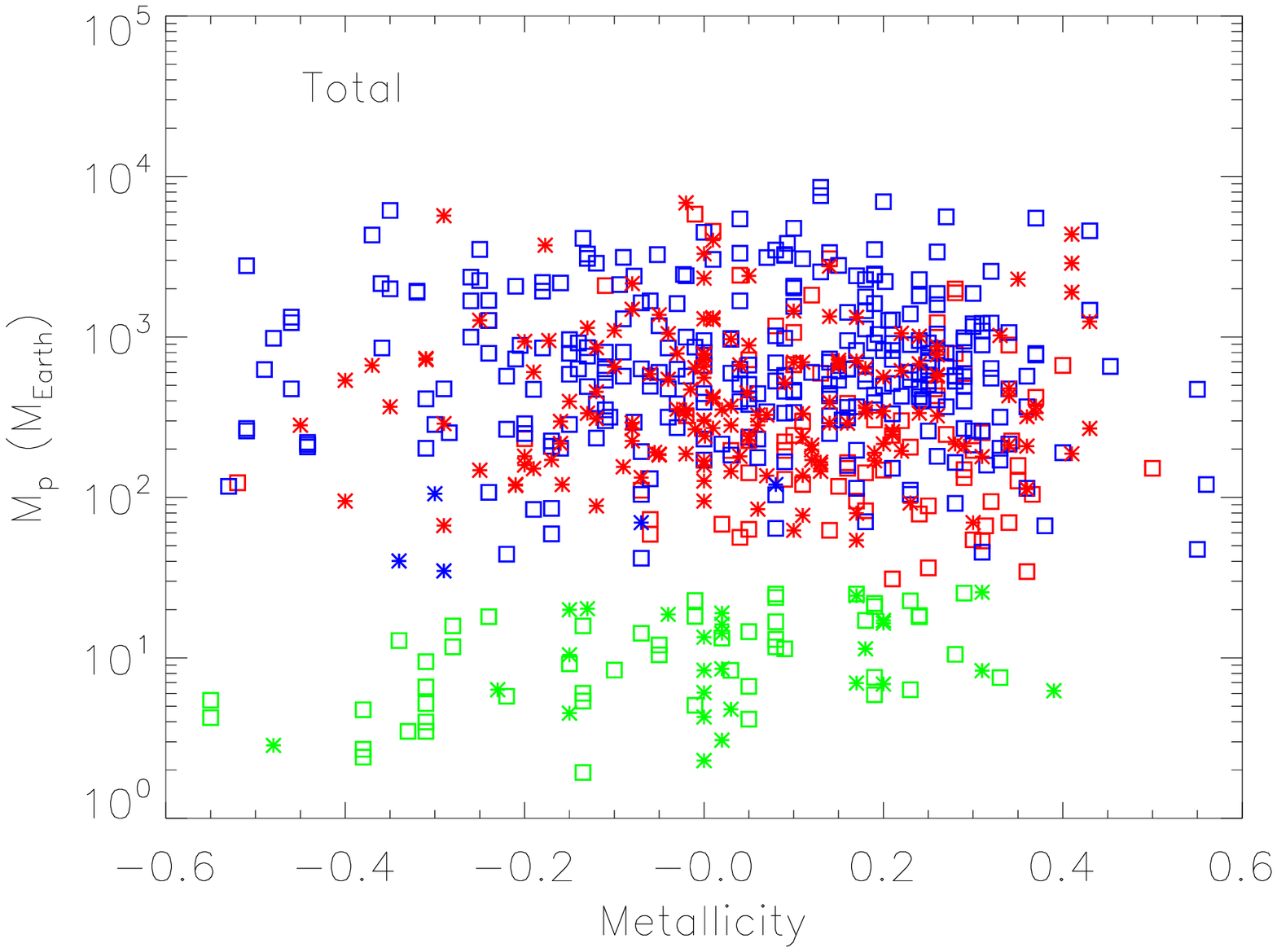}
\caption{Exoplanets observed by both radial velocity and transit methods. 
The mass of observed planets is shown as a function of the metallicity of the host stars. 
The data are taken from the Extrasolar Planets Encyclopaedia (http://exoplanet.eu/). 
Following the definition of the website, exoplanets detected first by the radial velocity technique are denoted by the squares 
whereas those first by the transit are by the stars. 
We note that the data are also used in HP13.
We divide these data for three zones: Hot Jupiters, Exo-Jupiters, and low-mass planets (see Table \ref{table1}). 
The observations show that both Jovian planets have a positive correlation with the stellar metallicity (see the two top panels) 
whereas low-mass planets are not correlated with it (the bottom left panel). 
Also, the RV data infer that the hot Jupiters may be slightly more sensitive to metallicity than the exo-Jupiters, 
although the trend may be weak (see the square). 
All the planets are shown in the bottom right panel.
Note that, although there may be some observational biases in the plot since the data are collected from a number of surveys, 
the trend of the planet-metallicity correlation is reproduced well.}
\label{fig1}
\end{center}
\end{minipage}
\end{figure*}

The rapid growth in the number of confirmed exoplanets ($\ga 1000$) and candidates ($\ga 3500$ from the {\it Kepler} mission) 
has opened several new frontiers in our understanding of planet formation  \citep{mq95,mb96,us07,mml11,brb12,fms14}.  
One of these is being driven by the wealth of information contained in the planet mass-semimajor orbital axis diagram 
wherein three distinct planetary populations emerge:  hot Jupiters in tight orbits $\simeq$ 0.1  AU, 
exo-Jupiters at intermediateorbital radii $\simeq$ 1  AU, and super-Earths between 1 and 10 Earth masses orbiting at small radii.  
Another,  which is the topic of this paper, 
is based on the correlation between exoplanet masses and the metallicity of their host stars. 
Gas giant planets (both the hot and exo-Jupiters) are observed with higher probabilities around stars with higher metallicities (high [Fe/H]$_*$), 
whereas there is little or no dependence on metallicity for low-mass planets such as super-Earths. 
The trend is known as the "planet-metallicity correlation",  and was originally suggested by radial velocity observations 
in the cornerstone paper by \citet{fv05} \citep[see also][]{g97,sim04,btb06}. 
In these observations, the metallicity distribution of target stars is approximately centered on [Fe/H]=0, 
so that the planet-metallicity correlation is not merely a consequence of the intrinsic distribution of stellar metallicities.
Recently, transit observations carried out by the {\it Kepler} mission have found a similar correlation 
\citep[although transit observations in general infer the size of planets, not the mass, e.g.][]{sl11,blj12}. 
Figure \ref{fig1} reproduces the trend.\footnote{
Note that the data of observed exoplanets are obtained from the Extrasolar Planets Encyclopaedia (http://exoplanet.eu/) 
in which the detections of a number of various surveys are combined.
As a result, there may be some biases in Figure \ref{fig1}, especially for the radial velocity data at the high metallicity regime.}

The planet-metallicity correlation is usually interpreted as evidence in favor of the core accretion scenario for the formation of Jovian planets. 
In this model, gas giants are built in two successive stages: 
the formation of rocky planetary cores by planetesimal collisions via both runaway and oligarchic growth \citep[e.g.,][]{ws89,ki98}, 
and the subsequent gas accretion onto these cores \citep[e.g.,][]{p96,lhdb09} 
if the mass of the planetary core grows to exceed a critical mass;  $M_{c,crit}$ \citep[e.g.,][]{m80,ine00}.  
The planet-metallicity correlation is a natural consequence of the core-accretion model 
because the final mass of cores via oligarchic growth increases with the solid density in protoplanetary disks \citep{ki02,tdl03} 
and the mass is the important parameter for initiating the subsequent gas accretion \citep{ine00,il04i}.
Population synthesis calculations based on this picture can generally reproduce the observational trend \citep{il04ii,mab12},
with most the observed gas giants predicted to form within $\sim$ 10 AU from 
their central stars. 

Further progress depends on the realization that  the value of the critical mass of planetary cores ($M_{c,crit}$) is of central importance to this theory. 
The canonical value for the core of Jovian planets at solar metallicity is  $ M_{c,crit} \simeq10 M_{\oplus}$ \citep{p96,ine00,lhdb09}.
Recent, more detailed studies however, have shown that $M_{c,crit}$ may be much smaller than 10 $M_{\oplus}$ \citep[e.g.,][]{hbl05,mbp10,hi11}. 
At lower metallicities, it is far from clear what its value should be. 
A larger value of  $M_{c,crit}$ with increasing metallicity would imply a metallicity-core mass correlation for massive planets. 
Low mass planets in our models may be regarded as "failed Jupiters" and arise by oligarchic growth \citep[see][]{hp13a} with
a small amount of gas accretion.
In order to explain the metallicity trend for low mass planets, 
the critical core mass would have to be relatively independent of metallicity in this regime. 
Do core accretion models indeed produce such a trend? 
What implications do the  observations have for such a model?  

In this paper, we apply our model for the formation and evolution of exoplanetary systems to this problem.
Our theory is based on the growth of cores at planet traps in evolving protoplanetary disks that 
we have developed in a series of earlier papers \citep[hereafter HP11, HP12, HP13]{hp11,hp12,hp13a}.
In our models, planet traps solve the basic problem of needing to drastically slow down the rapid type I migration of low mass planetary cores 
that is predicted in disks with smoothly varying density and temperature structure (HP11).  
The coupling of planet growth by core accretion to discrete, slowly moving traps (dead zones, ice lines, and heat transition radius) in inhomogenous disks 
gives rise to specific evolutionary tracks in the mass-semimajor axis diagram (HP12).  
The statistical scatter in the data and the existence of 3 planetary populations can be reproduced 
by a two parameter set of these evolutionary tracks (depending on distribution of disk lifetimes, and masses)  (HP13). 
In HP13, we developed a statistical tool, the planet formation frequency (PFF), 
to compute the percentage of evolutionary tracks from each of these three "feeder" traps, that contribute to each of the  3 planetary populations.   
We then showed that the model successfully explains the structure of the mass-semimajor axis diagram, 
in particular that exo-Jupiters at 1AU and super-Earth populations dominate at solar metallicity. 

Having demonstrated the consistency of our models with the populations in the mass-semimajor axis diagram,
we turn our attention here to the planet-metallicity diagram. 
Of particular importance for understanding this correlation is the behavior of the critical core mass ($M_{c,crit}$) as a function of metallicity.  
Given the large scatter in the data, it is unlikely that there is a single fixed number for  $M_{c,crit}$ at any metallicity.  
Our technique consists of adapting our statistical approach to evolutionary tracks
to compute the average value,  $\braket{M_{c,crit}} $  as a function of metallicity.   
We do this by computing a large number of evolutionary tracks and identifying the value $M_{c,crit}$ for each of these. 
The results depend upon the grain opacity in the planetary envelopes surrounding the cores,
and we use the data to constrain this typical value.  

Three important results emerge from our work.  
The first is that we find a planet-metallicity correlation for the cores, thereby pinning down the relation observed for exoplanets.  
The second is that we calculate the critical core masses for a wide range of metallicities of exoplanets. 
When applied to Jovian planets of solar metallicity, 
we infer that  $\braket{M_{c,crit}} \simeq 5 M_{\oplus}$ which has several interesting consequences for models of exoplanetary structure. 
And finally, we will show that low-mass planets are insensitive to metallicity. 
This is consistent with the observations of low-mass exoplanets, also known as super-Earths and/or hot Neptunes 
and implies that a large number of such observed planets can be formed as "failed" cores of gas giants and/or mini-gas giants.  

The plan of this paper is as follows. 
In Section \ref{core_mass}, we summarize the literature involved with the critical core mass as well as the associated gas accretion onto the cores.
In Section \ref{model}, we describe our semi-analytical model that is used for simulating planet formation and migration in gas disks, 
and discuss how the resultant PFFs are estimated, based on the model. 
In Section \ref{resu}, we present our results and discuss the statistical properties of planetary populations that are affected by metallicity. 
In Section \ref{para}, we perform a parameter study for examining how valid our findings are by varying key parameters. 
Also, we compare the results with the radial velocity observations 
and examine how useful our calculations are for putting valuable constraints on theoretical models. 
In Section \ref{disc}, we discuss a number of implications that can be derived from our results.  
Our conclusions are presented in $\S$ \ref{conc}. 

\begin{table*}
\begin{minipage}{17cm}
\begin{center}
\caption{Three zones for the observed exoplanets}
\label{table1}
\begin{tabular}{lccc} 
\hline 
Definition                   &  Mass range ($M_{\oplus}$)   &  Semimajor axis range (AU)  & HP13$^1$          \\ \hline 
                                 &                                                 &  0.01 $< r_p <$ 0.1                & Zone 1    \\ [-1ex]
\raisebox{1.5ex}{Hot Jupiters} & \raisebox{1.5ex}{30 $< M_p<$ $10^4$} & 0.1 $< r_p <$ 0.5 & Zone 2 \\[1ex] \hline
Exo-Jupiters              &  30 $< M_p<$ $10^4$            &  0.5   $< r_p <$ 10                 & Zone 3            \\[1ex] \hline
Low-mass planets    &  1 $< M_p<$ 30                      &  0.01   $< r_p <$ 0.5              & Zone 5           \\
\hline 
\end{tabular}

$^1$ the definitions adopted in HP13.
\end{center}
\end{minipage}
\end{table*}

\section{The critical core mass and subsequent gas accretion} \label{core_mass}

We turn first to a brief summary about the critical core mass as well as the 
relevant aspects of gas accretion that have been developed in the literature so far. 
We then present our prescription for them. 
The key quantities are summarized in Table \ref{table2}.

\subsection{Background}

The term "critical mass of planetary core" was coined in the early 1980's \citep[e.g.,][]{m80} in the context of the core accretion picture 
in which the formation of gas giants consists of two parts: 
rocky planetary cores form via collisional growth by planetesimals, followed by gas accretion onto the cores \citep[e.g.,][]{p96}.
At that time, {\it purely} hydrostatic calculations of gaseous envelopes around cores were performed which showed that 
hydrostatic envelopes cannot be maintained beyond the value of $M_{c,crit}$, 
and hence it was expected that efficient gas accretion onto the cores proceeds afterwords.  
The subsequent work in which {\it quasi-static} evolution of the gaseous envelopes were examined confirmed that 
rapid gas accretion takes place when $M_p > M_{c,crit}$ \citep{s82,bp86,ine00}.

The early calculations of this kind, by \citet{p96} constitute a milestone in this field.  Two new insights into the formation of gas giants were developed: 
1) rapid gas accretion occurs when the envelope mass becomes comparable to the core mass, (the mass was newly referred to as the crossover mass), 
and 2) the rapid gas accretion follows slow, gas accretion called  "phase 2". 
Phase 2 arises from the efficient accretion of planetesimals by cores. 
This is because the energy deposited by the  infall of planetesimals in the envelopes makes it possible 
to maintain the hydrostatic envelopes for a long time $\sim$ Myr 
(equivalently growing planetary cores satisfy the condition that $M_c \simeq M_{c,crit}$ 
due to a higher value of $\dot{M}_{c} (\sim 10^{-6} M_{\oplus} yr^{-1})$ (see Equation (\ref{m_ccrit})). 
This postpones the onset of rapid gas accretion. 
These two features were confirmed in several more recent calculations and simulations \citep{hbl05,lhdb09,mbp10}. 
Since the crossover mass physically corresponds to the critical core mass \citep{ine00,si08}, 
the results of Pollack et al therefore imply that the gas accretion can be determined largely by the behavior of the phase 2.

There is still some debate, however, about the existence of phase 2  \citep[e.g.,][]{ine00,fbb07}. 
For instance, \citet{si08} pointed out that phase 2 is unlikely to occur when the dynamics of planetesimals around a planet is considered more realistically. 
Specifically, they show through the numerical integration of planetesimals' orbit around a planet 
that a coupling effect of gravitational scattering by the planet and gas drag damping ends up with the formation of a gap in planetesimal disks. 
In other words, the efficient planetesimal accretion assumed in Pollack et al, to be the primary driver of the phase 2, is unlikely to be established.

When the planetesimal infall through the envelopes becomes negligible as the energy input, 
the gravitational contraction of the envelopes acts as the dominant heat source for supporting the envelopes. 
If this is the case, the growth of planetary mass via gas accretion can be regulated well by the Kelvin-Helmholtz timescales \citep[e.g.,][]{ine00,il04i}.

In summary, while significant progress has been achieved that has led to 
better understanding of the onset of gas accretion and the evolution of envelopes surrounding planetary cores,  
some physical processes remain uncertain.

\subsection{Our prescription \& important core masses}

Based on the above discussion, we adopt a conservative treatment
wherein the critical mass of planetary cores is used for initiating gas accretion onto the cores 
while the gas accretion is prescribed by the Kelvin-Helmholtz timescale \citep[HP12]{ine00,il04i}. 
One advantage of this formulation is that it allows us to compare our results
with those of the standard population synthesis calculations \citep[e.g.,][]{il04i,il08,iln13}, 
which incorporate such an approach.  

The critical core mass ($M_{c,crit}$) depends on the rate of planetesimal accretion onto the core ($ \dot{ M}_c $) 
and the grain opacity ($ \kappa $) in envelopes around the cores  \citep[e.g.,][]{ine00,il04i,hi11}.  This was written in HP12 as
\begin{equation}
 M_{c,crit} \simeq 10 M_{\oplus} f_{c,crit} \left( \frac{\dot{M}_c}{10^{-6}M_{\oplus} \mbox{ yr}^{-1}} \right)^{1/4},
 \label{m_ccrit}
\end{equation}
where  $f_{c,crit}$ is treated as a free parameter. 
This parameter is directly related to the grain opacity in the envelopes \citep{ine00,il04i}; $f_{c,crit} = (\kappa / 1\mbox{ cm}^2 \mbox{ g}^{-1})^{0.2-0.3}$  
whose variation is usually neglected.  
Many previous studies adopt $f_{c,crit}= 1$ as the canonical value in which both the abundance and the size distribution of 
dust grains in the envelopes are assumed to be similar to the interstellar medium (ISM) \citep{p96,ine00,il04i}. 
In our study, the value of $f_{c,crit}$ plays an important role as we investigate systems with lower metallicity, and we demonstrate 
later that a much lower value fits the extrasolar planet data better.  

For notational convenience, we define and shall henceforth adopt a mass parameter that  
arises from the dependence of $M_{c,crit}$ on the grain opacity, and that links to definitions used in our previous work, namely;
\begin{equation}
\label{m_ccrit0}
M_{c,crit0} \equiv 10 M_{\oplus} f_{c,crit} = 10 M_{\oplus} \left( \frac{\kappa} {1 \mbox{ cm}^2 \mbox{ g}^{-1}} \right)^{0.2-0.3}.
\end{equation} 

One might think  that the isolation mass of cores ($M_{c,iso}$) which is the calculated core mass, 
assuming that all the planetesimals in the feeding zone are accreted onto the cores (see Equation (\ref{mc_iso})),
would be better
for examining the planet-metallicity correlation.
However, the value of $M_{c,iso}$ is really useful only when the formation of cores occurs in-situ.
In the case of migrating cores - in particular those caught in our moving planet traps -  
fresh planetesimals can always be introduced as the trap sweeps through the 
disk.   It is not so obvious that an isolation mass is the correct choice in this situation.   Hence  $M_{c,crit}$ is preferred in our models.

Gas accretion onto the cores starts once they achieve the value of $M_{c,crit}$ \citep[e.g.,][]{m80,s82,ine00}. 
The gaseous envelopes of potential Jovian planets contract on the Kelvin-Helmholtz timescale, 
provided that the accretion of planetesimals by the cores is small ($< 10^{-6} M_{\oplus} yr^{-1}$) at that time.  
A simplified version of this timescale has been derived from more detailed numerical simulations \citep{ine00}:
\begin{equation}
 \tau_{KH} \simeq 10^{c} \mbox{ yr} \left( \frac{M_p}{M_{\oplus}} \right)^{-d}.
 \label{tau_KH}
\end{equation}
It is likely that there are ranges in $c$ ($8 \la c \la 10$) and in $d$ ($2 \la d \la 4$). 
We adopt that $c=9$ and $d=3$ as fiducial values, following \citet[references herein]{il04i}.
Note that $\tau_{KH}$ also depends on the grain opacity, $\kappa$, like $M_{c,crit0}$ (see Equation (\ref{m_ccrit0})). 
Nonetheless, the dependency is neglected in this paper, because it can be incorporated into the variation of $c$, and $d$, 
both of which are the main parameters here.
Also, note that $\tau_{KH}$ is sensitive to the computed mass ($M_p$) of planets. 
For instance, $\tau_{KH} = 8 \times 10^6$ yr when $M_p=5M_{\oplus}$, 
which roughly corresponds to the onset of gas accretion for our fiducial case (see below). 
When $M_p =10 M_{\oplus}$ the timescale goes down to $10^6$ yr. 
Our analysis will show that the lower critical core mass parameter fits the data
better, which means that this model mimics the behavior of a drawn out phase 2 found by \citet{p96} but whose physical origin is different: 
for our case, a prolonged phase 2-like behaviour originates simply from a longer Kelvin-Helmholtz timescale, and not with planetesimal accretion.

The gas accretion rate onto the protoplanet from the surrounding disk is regulated by this Kelvin-Helmholtz timescale; 
\begin{equation}
 \frac{d M_p}{dt} \simeq \frac{M_p}{\tau_{KH}} \propto M_p^{d+1}.
 \label{gas_acc}
\end{equation}
Thus, the growth rate of planets increases sharply with $M_p$. 
 
There is one final mass that needs to be considered in this accretion picture, which is the mass scale at which the gas accretion finally runs away. 
The minimum possible mass for a core is $M_{c,min}$ that can be derived from the condition 
$\tau_{KH}=\tau_{fin}$ (see equation (\ref{tau_KH})). 
Setting $\tau_{fin}=10^7$ yr, which is the upper limit of the disk lifetime, we obtain
\begin{equation}
\label{m_cmin}
M_{c,min}=10^{(c-7)/d} M_{\oplus}.
\end{equation}
For our fiducial case  ($c=9$ and $d=3$), we find  $M_{c,min} \simeq 4.6 M_{\oplus}$.

To summarize, gas accretion starts when $M_c \geq M_{c,crit}$ (see Equation (\ref{m_ccrit})), 
and the subsequent growth of planetary mass via the gas accretion is determined by the Kelvin-Helmholtz timescale 
(see Equations (\ref{tau_KH}) and (\ref{gas_acc})). 
These prescriptions involve the three key physical quantities 
that are basic control parameters in our models;
$M_{c,crit0}$, $c$, and $d$ (see Table \ref{table2}). 
The fundamental question addressed by this paper is how 
the value of the grain opacity defining $M_{c,crit0}$ (see equation (\ref{m_ccrit0})), 
affects the PFFs and so determines the planet-metallicity correlation (see Section \ref{resu}). 
In pursuing this question, we investigate the effects of $c$ and $d$ (see equation (\ref{tau_KH})) in Section \ref{para} (see Table \ref{table4}).

\begin{table*}
\begin{minipage}{17cm}
\begin{center}
\caption{Key quantities in this work}
\label{table2}
\begin{tabular}{lll}
\hline
Symbol             &  Meaning                                                                                                                                     & Values           \\ \hline
                         &   Planetary growth by gas accretion                                                                                           &                                       \\ \hline
$M_p$              &   Planetary mass (the core mass + the envelope mass) computed along tracks                                        &         \\
$M_{c,crit}$      &   The critical mass of planetary cores that can initiate efficient gas accretion (see Equation (\ref{m_ccrit}))   &                \\
$\dot{M}_c$      &   Accretion rate of planetesimals by planetary cores (see Equation (\ref{mc_dot}))                        &                   \\
$M_{c,crit0}$    &   A free parameter regulating $M_{c,crit}$ (see Equations (\ref{m_ccrit}) and (\ref{m_ccrit0}))      &  $3M_{\oplus}$, $5M_{\oplus}$, $10M_{\oplus}$                \\
$\kappa$          &   The grain opacity in accreting envelopes that links to $M_{c,crit0}$ (see Equation (\ref{m_ccrit0}))   & $\simeq$8 $\times 10^{-3}$, 0.06, 1 cm$^2$ g$^{-1}$                  \\   
$\tau_{KH}$       & The Kelvin-Helmholtz timescale (see Equation (\ref{tau_KH}))                                                     &                                                \\   
$(c,d)$               & a set of free parameters for $\tau_{KH}$                                                                                        & (9,3)                                        \\
$M_{c,min}$      &  The mass scale for gas accretion to finally run away within the disk lifetime (see Equation (\ref{m_cmin}))  &  $\simeq 4.6M_{\oplus}$  \\ \hline
                         &  Disk model                                                                                                                                  &                                       \\ \hline
$\dot{M}$          &  Disk accretion rate onto the central star  (see Equation (\ref{mdot_exp}))                                     &               \\
$\Sigma_g$      &  The surface density of gas ($\propto \dot{M}$)                                                                           &               \\
$\Sigma_d$      &  The surface density of dust ($=f_{dtg} \Sigma_g$)                                                                       &               \\
$\mbox{[Fe/H]} $ &  Metallicity (see Equation (\ref{z_disk}))                                                                                     &               \\
$f_{dtg}$           &   the dust-to-gas ratio ($=f_{dtg, \odot} \eta_{dtg}$)                                                                       &                                    \\
$f_{dtg, \odot}$ &   the dust-to-gas ratio at the solar metallicity  ($=\eta_{dtg, \odot} \eta_{ice}$)                             & $\simeq 1.8\times10^{-2}$      \\
$\eta_{dtg, \odot}$ &  the dust-to-gas ratio at the solar metallicity within the ice line                                             & $\simeq 6\times10^{-3}$   \\
$\eta_{ice}$      &   A factor for increasing/decreasing $f_{dtg, \odot}$ due to the presence of ice lines                            &                                      \\
$\tau_{vis}$      &   the viscous timescale (see Equation (\ref{mdot_exp}))                                                                 &   $10^6$ yr                  \\               
$\tau_{int}$      &   the initial time for starting computations (see Equation (\ref{mdot_exp}))                                    &   $10^5$ yr                  \\                       
\hline
\end{tabular}
Note that free parameters here are only $M_{c,crit0}$, $c$, and $d$. 
For the other parameters,  the given values are derived from the three free parameters, 
or our previous study confirmed that the results are insensitive to a specific choice of the values (HP13).
\end{center}
\end{minipage}
\end{table*}

\section{Statistical treatment} \label{model}

We describe our statistical modeling that utilizes the semi-analytical model developed in a series of our papers (HP11, HP12, HP13). 
We briefly discuss our approach and refer the readers to these three papers for further details (see Table \ref{table2}, also see Appendix \ref{app1}).

\subsection{The data}

We obtained the data given in Figure 1 from the Extrasolar Planets Encyclopaedia \citep[http://exoplanet.eu/]{sdl11} 
and plot the mass of observed exoplanets as a function of metallicity for the 3 different populations in Figure \ref{fig1}.  
Table \ref{table1} summarizes the definition of these 3 populations.
The identification of different exoplanetary populations was discussed in earlier work \citep[HP13]{cl13}. 
Since our previous study already confirmed 
that the population of close-in ($r \leq 0.5$ AU) massive planets (the populations ending up either in Zone 1 or in Zone 2) are  minor (HP13), 
we combine these two populations in this paper and refer to them as "hot Jupiters" (see Table \ref{table1}).
The classification of Jovian planets does not affect our conclusions.
 
Figure \ref{fig1} clearly shows that the number of more massive planets gradually diminishes as one goes to lower metallicity, 
although there is a large scatter.
For the low mass planets, on the other hand, this trend is absent.  
These results were initially inferred from the radial velocity observations \citep[e.g.,][]{sim04,fv05,btb06,us07}. 
Note that some observational biases may be present in Figure \ref{fig1}, 
since the observational data are obtained from the collection of various surveys.
Nonetheless, the trend of the planet-metallicity correlation is evidently shown.
Also note that planets with measured masses (the radial velocity method) as well as those inferred from the Kepler observations 
show similar patterns with metallicity.   

\subsection{Basic model}

\begin{figure}
\begin{center}
\includegraphics[width=9cm]{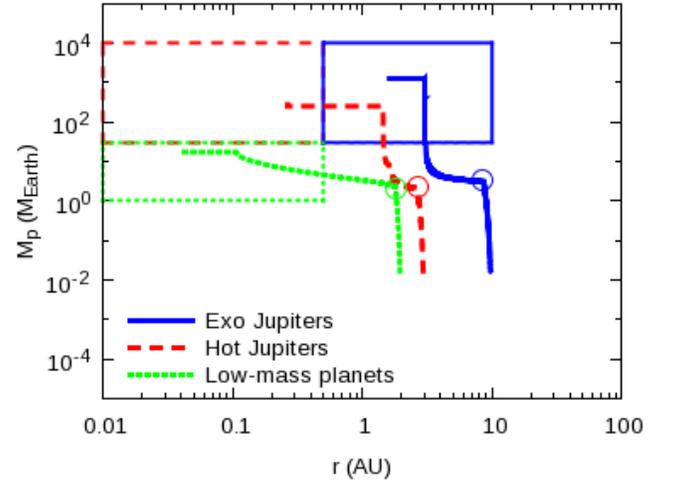}
\caption{Examples of evolutionary tracks computed in the mass-semimajor axis diagram. 
Different tracks end up in different zones. 
The circles denote the values of $M_{c,crit}$ for these examples.}
\label{fig2}
\end{center}
\end{figure}

We constructed evolutionary tracks of accreting planets in the mass-semimajor axis diagram
by computing the time evolution of viscous disks with photoevaporation of gas in HP12. 
For the growth of planetary cores, we adopted the standard core accretion scenario \citep[HP12,][also see Appendix \ref{app1}]{il04i} 
in which rocky planetary cores form via oligarchic growth \citep{ki98}. 
Our prescription for the subsequent growth of planets by gas accretion was discussed in Section \ref{core_mass}.
In Figure \ref{fig2}, we give examples of 3 computed evolutionary tracks that feed the different populations in the mass-semimajor axis diagram. 

Planets are built  in planet traps which naturally arise from some kinds of disk inhomogeneities.
As the term implies, they act as barriers (technically as radii at which there is zero-net torque exerted by the disk) 
for rapid type I planetary migration \citep[HP11]{mmcf06,mpt07,il08v,hp10,lpm10}. 
We considered three types of planet traps: dead zones, ice lines and heat transition traps (HP11). 
The orbital evolution of planets forming in them is regulated by the movement of planet traps and subsequent type II migration. 
When planetary cores become more massive than the gap-opening mass, 
they drop out of their traps and switch to type II migration.
Although it can be assumed that planet formation proceeds predominantly at three types of traps treated in isolation, 
the gravitational interactions between trapped planets will gradually become important for the later stage of disk evolution, 
an effect that is neglected
in this work but will be treated in a future paper.

In building a theoretical evolutionary track (see HP12), we first compute the growth of the rocky 
core from the theoretical expression for $\dot{M_{c}}$ on the core accretion timescale
$\tau_{c,acc}$ having to do with the collisions of planetesimals.  
One integrates the mass over time taking into account the fact that the accreting core is moving through the disk as it is carried by its trap. 
The actual formula for $\tau_{c,acc}$ and $\dot{M}_c$ are given by Equations (\ref{tau_cacc}) and (\ref{dotM_c}) in Appendix \ref{app1} and 
follows \citet[also see HP12]{ki02}.  
In this regime, the accretion rate is directly proportional to the surface density of the dust; $ \dot{M_c} \propto \Sigma_d$, 
which as we will see in the next subsection, is related to the metallicity of the disk.  
This column density changes as the host planet trap moves radially inwards through the evolving disk so that $ \dot{M_c}$ changes with time.   
At the time in the numerical integration when the mass of the core exceeds the critical value given by condition (see equation (\ref{m_ccrit})), 
one has identified the critical core mass ($M_{c,crit}$).  

In Figure \ref{fig2}, we mark the value of $M_{c,crit}$ found for each of the tracks.  
Beyond this mass, the evolution of the planet is switched to gas accretion which 
is  governed by the Kelvin-Helmoltz timescale given above (see equations (\ref{tau_KH}) and (\ref{gas_acc})).    
As discussed in Section \ref{core_mass}, in order to achieve rapid gas accretion, 
one needs to have $M_{c,crit} \ge M_{c,min}$ (see equation (\ref{m_cmin})). 

The protoplanet drops out of the trap motion and switches to subsequent type II migration 
when it becomes sufficiently massive to open a gap in the disk.  
We can see from Figure 2 that tracks that end up in the hot Jupiter zone must undergo slower gas accretion overall 
if they are to end up at such small orbital radii.   
This is a manifestation of type II migration where gap opening will significantly reduce the gas inflow onto the core.  
Tracks that end up in the exo-Jupiter zone are therefore less dominated by type II gap-opening effects, 
and hence accrete more rapidly to achieve higher masses.   
The final phase in the evolution arrives with the dissipation of the disk by photoevaporation.      
 
We emphasize that all of this planetary evolution and movement occurs in disks whose properties are evolving at the same time, 
due to the viscous dissipation of disks as well as their relatively rapid dispersal by photoevaporation characterizing the final stage of disk evolution.  
In order to make this quantitative, we have focused on two key disk parameters: the disk accretion rate and lifetime 
parameters.\footnote{In fact, there are three important parameters, the disk mass, lifetime, and accretion rate. 
As discussed in HP13, however, these three are linked together, so that we pick up the disk accretion rate and lifetime as the main parameters.} 
In our improved formalism, these two parameters involve the time-evolution of the disk accretion rate that is given as 
(HP13, also see Tables \ref{table2} and \ref{table3})
\begin{eqnarray}
 \label{mdot_exp}
 \dot{M}(\tau) & \simeq & 3 \times 10^{-8} M_{\odot} \mbox{ yr}^{-1} \eta_{acc} 
                                                   \left( \frac{M_*}{0.5M_{\odot}}\right)^2 \\
              &    &  \times \left(1+ \frac{\tau}{ \tau_{vis}} \right)^{-1.5}  \exp \left( - \frac{\tau-\tau_{int}}{\tau_{dep}}  \right),  \nonumber                                   
\end{eqnarray}
where $M_*=1 M_{\odot}$ is the stellar mass, $\tau$ is a time, $\tau_{vis}=10^6$ yr is the typical viscous timescale, 
$\tau_{dep}$ is the depletion timescale, and $\tau_{int}=10^5$ yr is the initial time for our calculations to get started. 
The numerical factor in equation (\ref{mdot_exp}) is set so as to
well match the observed disk accretion rate for disks around classical T Tauri stars ($M_* \simeq 0.5 M_{\odot}$) 
at $\tau=10^6$ yr, if a factor $\eta_{acc}$ is unity. 
Since the total disk mass is linearly proportional to $\dot{M}$ in our model, we can essentially change the disk mass by varying $\eta_{acc}$. 
We also assume that the depletion timescale can be written as 
\begin{equation}
\label{tau_dep}
\tau_{dep} = \tau_{dep,0} \eta_{dep}, 
\end{equation}
where $\tau_{dep,0}=10^6$ yr. 

In our formalism, the disk evolution is therefore regulated by two quantities, $\tau_{vis}$, and $\tau_{dep}$. 
The disk lifetime is controlled largely by $\tau_{dep}$. 
Following HP13, we adopt the exponential function for characterizing the end disk evolution stage. 
As discussed in HP13 (see its Appendix), the exact shape of the disk accretion, especially at the end stage of disk evolution, 
does not affect the resultant planetary populations (see equation (\ref{pfr})). 
The most important disk parameter is the disk lifetime. 
In summary, we treat $\eta_{acc}$ as a disk accretion rate parameter whereas $\eta_{dep}$ is considered as a disk lifetime parameter.

Our model (HP12) succeeded in providing physical explanations for a number of observational features 
such as the observed mass-period relation \citep{us07,mml11}. 

\subsection{Metallicities}

The metallicity of the disk is readily parameterized in terms of the dust-to-gas ratio $f_{dtg}$.   
This can be written as $f_{dtg}=f_{dtg,\odot} \eta_{dtg} $ (see Tables \ref{table2} and \ref{table3}), 
where $f_{dtg,\odot}=\eta_{dtg,\odot} \eta_{ice}$ is the dust-to-gas ratio at the solar metallicity, 
$\eta_{dtg}$ is a factor for varying $f_{dtg}$, $\eta_{dtg,\odot}$ is the the dust-to-gas ratio at the solar metallicity within the ice line, 
and $\eta_{ice}$ is a factor for changing $f_{dtg, \odot}$ due to the presence of a water ice line. 
We set $\eta_{ice}=1$ when planet traps are inside the ice line, $\eta_{ice}=4$ when the traps are on the ice line, 
and $\eta_{ice}=3$ when the traps are beyond the ice line \citep[HP13]{phbs94,il04i}. 
Adopting the most likely value of $f_{dtg,\odot}$, which is $\sim 1.8 \times 10^{-2}$ \citep{ags09}, we obtain that $\eta_{dtg,\odot}=6 \times 10^{-3}$. 
Thus, the variation of the dust-to-gas ratio is regulated by $\eta_{dtg}$. 

In our model, $\eta_{dtg}$ is related to the dust density in disks: 
 $\Sigma_d = f_{dtg} \Sigma_g (\propto \eta_{dtg} \dot{M})$, where $\Sigma_g (\propto \dot{M})$ is the surface density of gas in the disks.
As shown by the previous studies \citep[e.g.,][]{il04ii,mab12}, 
the efficiency of forming planetary cores in the core accretion scenario is sensitive to the value of $\Sigma_d$.

Finally,  the metallicity can be expressed in terms of $\eta_{dtg}$. 
Assuming that the value of the metallicity in disks is similar to their stellar metallicities, 
the metallicity [Fe/H]\footnote{We adopt the conventional definition, 
so that [Fe/H] is the metallicity measured from the solar metallicity, that is, [Fe/H]=0 for the solar metallicity.} 
can be linked with $\eta_{dtg}$ as follows \citep{il04ii}:

\begin{equation}
 \label{z_disk}
\mbox{[Fe/H]}  \simeq \mbox{ [m/H]} \equiv \log_{10} \frac{f_{dtg}}{f_{dtg,\odot}} = \log_{10}(\eta_{dtg}),
\end{equation}
where m represents a mixture of metals. This indicates that the effects of metallicities can be readily explored by varying $\eta_{dtg}$. 
Note that the ratio of [m/H] to [Fe/H] is generally considered as order of unity.

\subsection{Initial conditions}

We adopt the same initial conditions as HP13 (also see HP12), 
wherein protoplanets of $\sim$0.01 $M_{\oplus}$ start to grow at specific times $\tau$ and positions $r$. 
We take 100 times (and 100 positions) for each planet traps as the initial conditions of evolutionary tracks, 
so that the entire disk lifetime is covered by the tracks. 
This is equivalently that $N_{int}=300$ for all the calculations (see equation (\ref{pfr})). As confirmed in HP13, 
the number is large enough to get our results convergent.

For the value of the metallicity ([Fe/H]), we consider a wide range ($-0.6 \leq$ [Fe/H] $ \leq 0.6$). 
The variation in [Fe/H] leads to the change in $\eta_{dtg}$ through equation (\ref{z_disk}). 
Utilizing the above model with these initial conditions, we evaluate the PFFs for three zones as a function of metallicity.

\subsection{Planet formation frequencies}

In  HP13, we developed a new statistical approach for computing the contributions of the various planet traps to the three major planetary populations 
that does not use a standard population synthesis approach.  
We  partitioned the mass-semimajor axis diagram into 5 zones and calibrated planet formation frequencies (PFFs) in the zones.  
Here, we consider three zones which characterize the dominant populations (see Table \ref{table1} and Figure \ref{fig2}, also see HP13). 
The input parameters for the statistical analysis are summarized in Table \ref{table3}.

\begin{table*}
\begin{minipage}{17cm}
\begin{center}
\caption{Input parameters for the statistical analysis$^1$}
\label{table3}
\begin{tabular}{ccc}
\hline
Symbol             &  Meaning                                                                                                                                     & Values           \\ \hline
                         &  Stellar parameters                                                                                                                      &                                       \\ \hline
$M_*$               &  Stellar mass                                                                                                                                & 1 $M_{\odot}$               \\
$R_*$               &  Stellar radius                                                                                                                                &  1 $R_{\odot}$              \\
$T_*$               &  Stellar effective temperature                                                                                                        & 5780 K                           \\ \hline
                         &  Disk mass parameters                                                                                                                 &                                       \\ \hline
$\eta_{acc}$     &  A dimensionless factor for $\dot{M}$ (see Equation (\ref{mdot_exp}))                                         & $0.1 \leq \eta_{acc} \leq 10$  \\
$w_{mass}(\eta_{acc})$ & Weight function for $\eta_{acc}$ modeled by the Gaussian function                                &                                    \\
$\eta_{dtg}$      &   A parameter for increasing/decreasing  [Fe/H] (see Equation (\ref{z_disk}))                              & $0.25 \la \eta_{dtg} \la 4$  \\ \hline
                         &  Disk lifetime parameters                                                                                                               &                                      \\ \hline   
$\eta_{dep}$       &   A dimensionless factor for  $\tau_{dep}$ (see Equation (\ref{tau_dep}))                                    &  $0.1 \leq \eta_{dep} \leq 10$      \\                                                                                                                                              
$w_{lifetime}(\eta_{dep})$ & Weight function for $\eta_{dep}$ modeled by the Gaussian function                           &                                    \\
\hline
\end{tabular}

$^1$ These quantities are constrained by the observations, so that they are not free parameters.
\end{center}
\end{minipage}
\end{table*}

Utilizing the calculations of evolutionary tracks of planets building at planet traps (HP12), we define the PFFs for
tracks feeding into the ith zone as (HP13)
\begin{eqnarray}
 \label{pfr}
 \mbox{PFFs(Zone i)}   & \equiv  &  \\  \nonumber   
           \sum_{\eta_{acc}} \sum_{\eta_{dep}} & w_{mass}(\eta_{acc}) & w_{lifetime}(\eta_{dep}) \\  \nonumber        
          & \times   & \frac{N\mbox{(Zone i, } \eta_{acc}, \eta_{dep})}{N_{int}} \nonumber                           
\end{eqnarray}
where $ w_{mass}$ and $w_{lifetime}$ are both weight functions of $\eta_{acc}$ and $\eta_{dep}$, respectively, 
$N\mbox{(Zone i, } \eta_{acc}, \eta_{dep})$ is the number of evolutionary tracks that end up in Zone i for a given set of $\eta_{acc}$ and $\eta_{dep}$, 
and $N_{int}=300$ is the total number of tracks that are considered in the calculations. 

Briefly,  each term in the weighted sum on the right hand side of equation (\ref{pfr}) involves the ratio, $N\mbox{(Zone i, } \eta_{acc}, \eta_{dep})/N_{int}$, which
is  a planet formation {\it efficiency} for a specific set of $\eta_{acc}$ and $\eta_{dep}$. 
This ratio is then summed (integrated in the continuum limit)  over a wide range of both $\eta_{acc}$ and $\eta_{dep}$ with their weight functions to produce the PFF.  
Therefore the PFF  represents the (integrated) planet formation efficiencies for a wide range of disk mass and lifetime. 
In this paper, we consider the ranges of $\eta_{acc}$ ($0.1 \le \eta_{acc} \le 10$) and $\eta_{dep}$ ($0.1 \le \eta_{dep }\le 10$), following HP13.

We adopt Gaussian distributions for both $w_{mass}$ and $w_{lifetime}$. 
For $w_{mass}$, the peak value is obtained when $\dot{M} \simeq 1.7 \times 10^{-8} M_{\odot} yr^{-1}$ (equivalently $\eta_{acc}=1$). 
The standard deviation of $w_{mass}$ is set as unity in the unit of $M_{\odot} yr^{-1}$, 
although we confirmed that its variation does not affect our results. 
The choice of these particular values is  motivated by the observed disk accretion rate around classical T Tauri stars. 
For $w_{lifetime}$, the highest value is achieved when the disk lifetime is about $1.5 \times 10^6$ yr (equivalently $\eta_{dep}=1.5$) 
with the standard deviation being 3 in the unit of Myr. 
These two values are again selected to be consistent with the observations of disks 
which show that a disk fraction of stars can be well fitted by the exponential function with the e-folding time being 2.5 Myr \citep[e.g.,][]{wc11}. 
Thus, all the quantities in the weight functions are determined, so that the weight functions can fit to the observations of protoplanetary disks (see HP13). 
Gaussian functions are also useful, both because of their computational convenience as well
as their reflection of possible Gaussian random noise in the initial conditions for disk formation in turbulent molecular clouds.
 
This approach enables one to compare the theoretical results with the observations without the standard population synthesis calculations being performed.
As discussed in HP13, we have shown that 
the combination of planet traps with the core accretion scenario can account for the observations of exoplanets very well.  
Specifically, we have demonstrated that most formed gas giants tend to end up around 1 AU with fewer populations of the hot Jupiters. 
Also, our calculations have shown that a large fraction of low-mass planets in tight orbits can be populated by "failed" cores of gas giants and/or mini-gas giants. 

\subsection{A statistical approach for critical core masses}

We adapt our PFF approach to calculating the  average critical mass of cores as a function of metallicity for a distribution of evolutionary tracks.  
In order to proceed, we calculate the mean mass of planetary cores ($\braket{M_{c,crit}}$) that indeed initiate gas accretion 
when evolutionary tacks are computed, which is defined as
\begin{eqnarray}
\label{M_c_mean}
\braket{M_{c,crit}(\mbox{Zone i})} & \equiv & \\  \nonumber   
  \sum_{\eta_{acc}} \sum_{\eta_{dep}} & w_{mass}(\eta_{acc}) &  w_{lifetime}(\eta_{dep}) \\  \nonumber    
                                                            & \times &   \braket{M_{c,crit}(\mbox{Zone i, } \eta_{acc}, \eta_{dep})},
\end{eqnarray}
where $\braket{M_{c,crit}(\mbox{Zone i, } \eta_{acc}, \eta_{dep})}$ is the averaged critical mass of planetary cores that eventually fill out Zone i and that 
just start accreting gas during planetary growth for a certain set of $\eta_{acc}$ and $\eta_{dep}$.  
We have shown in Figure \ref{fig2} how these values are determined for each evolutionary track computed in our simulations (see the circle in each track).  

\section{Results} \label{resu}

\begin{figure*}
\begin{minipage}{17cm}
\begin{center}
\includegraphics[width=8cm]{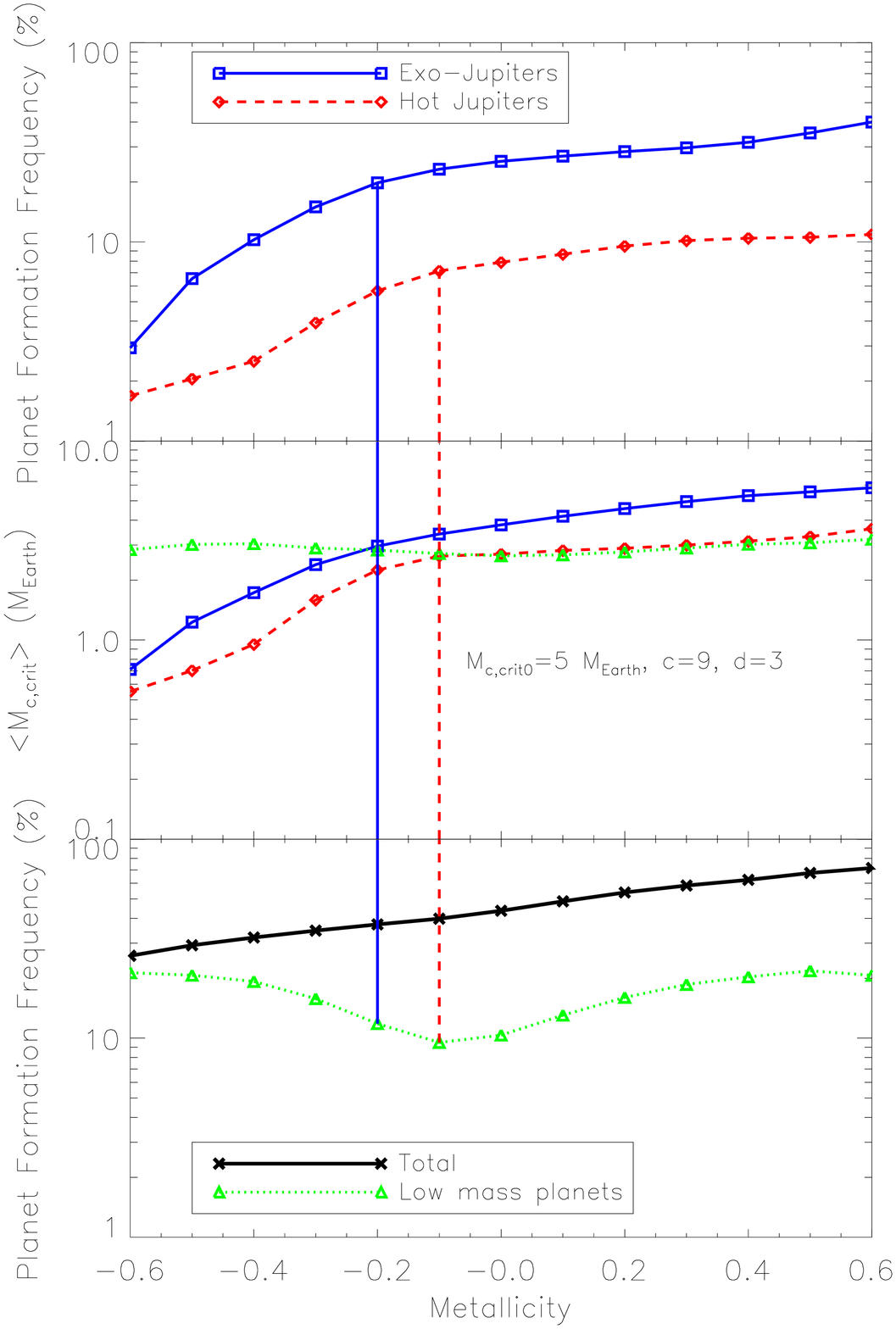}
\includegraphics[width=8cm]{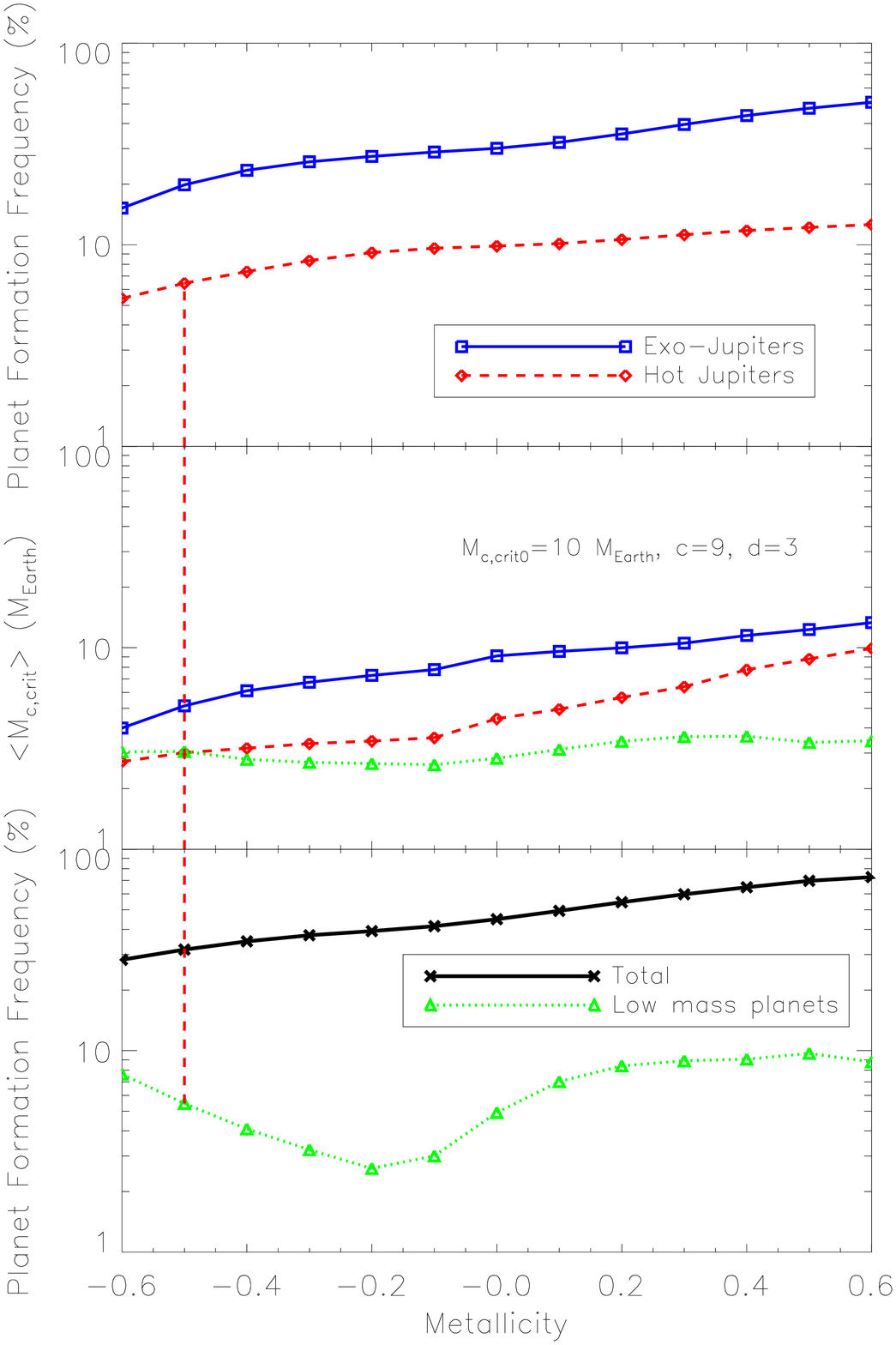}
\caption{The resultant PFFs and the mean mass of planetary cores as a function of the metallicity (see Table \ref{table4}). 
The results for the case of $M_{c,crit0}=5M_{\oplus}$ and $M_{c,crit0}=10M_{\oplus}$ are shown in the left and right panels, respectively. 
The top panels show the PFFs of both the hot and exo-Jupiters whereas the bottom panels show the total PFFs and those of low mass planets. 
The middle panels show the mean critical mass of planetary cores ($\braket{M_{c,crit}}$) that indeed start accreting gas in our model (see equation (\ref{m_ccrit})). 
The results show that both the hot and exo-Jupiters are readily formed in high metallicity disks and their PFFs decrease for low metallicity disks 
(see the dashed and solid lines for the hot and exo-Jupiters, respectively). 
On the contrary, the PFFs of low-mass planets do not decrease steadily for disks with lowering metallicities (see the dotted line). 
The total PFF is a steadily increasing function of the metallicity (see the thick line). 
We find that the intersections between the values of $\braket{M_{c,crit}}$ for Jovian planets and low mass planets 
enable us to estimate transition metallicities (TMs) below which the PFFs for Jovian planets drop suddenly 
(see the vertical dashed and solid lines for the hot and exo-Jupiters, respectively). 
It is important that the results for the case of $M_{c,crit0}=5M_{\oplus}$ are more likely to match the current available observations of exoplanets.}
\label{fig3}
\end{center}
\end{minipage}
\end{figure*}

We computed complete sets of evolutionary tracks for three different values of the $M_{c,crit0}$ parameter; 
($3M_{\oplus}, 5M_{\oplus}$ and $10M_{\oplus}$) for the range of metallicities seen in the data (see Table \ref{table4}). 
We calculate, in particular, the PFFs as a function of metallicity for the 3 different planetary populations, 
and discuss how different the resultant PFFs are for different $M_{c,crit0}$. 
We also determine the values of the critical core masses ($\braket{M_{c,crit}}$) as a function of metallicity, 

\subsection{PFFs for Jovian planets}

Figure \ref{fig3} (top) shows the computed PFFs as a function of metallicity for the hot and exo-Jupiters 
(see the dashed and solid lines, respectively). 
The results for the choices $M_{c,crit0}=5M_{\oplus}$ and $M_{c,crit0}=10M_{\oplus}$ are shown on the left and right panels, respectively. 
For both values of $M_{c,crit0}$, we find that the PFFs of the exo-Jupiters are much higher than those of the hot Jupiters. 
Thus, the results found in HP13 that exo-Jupiters should be much more frequent than hot Jupiters 
carries over to planets of all observed metallicities.  
We predict therefore that a higher population of Jovian planets around 1AU is general, 
and is a reflection of the dynamics of planet traps in the metallicity range so far observed.  

Our second major result apparent in this figure is that the formation of both Jovian planets is sensitive to the metallicity: 
gas giants are preferentially formed for disks with higher metallicities whereas their formation efficiencies are suppressed for low metallicity disks. 
This agrees very well with the sense of the observed planet-metallicity relation - 
massive planets are less frequently observed around stars of lower metallicity.  
This is a natural consequence of planet formation in the core-accretion picture in our calculations.
These results also agree with the findings of previous population synthesis studies \citep{il04ii,mab12}.

A very interesting and new result is that the PFFs  drop off very noticeably towards lower metallicities 
and remain rather constant above some particular values.  
For the case of $M_{c,crit0}=5M_{\oplus}$ (the left panel), the PFFs of both the hot and exo-Jupiters drop rapidly at low metallicity, [Fe/H]= -0.2 to -0.4. 
For the case of $M_{c,crit0}=10M_{\oplus}$ (the right panel), on the contrary, the PFFs of both of them decrease more gradually as the metallicity decreases.  
We explore this further in what follows.  

\begin{table*}
\begin{minipage}{17cm}
\begin{center}
\caption{Summary of the calculations and their results}
\label{table4}
\begin{tabular}{ccccclcc}
\hline
                 & $M_{c,crit0}$ ($M_{\oplus}$)  &  $c$  &  $d$  & $M_{c,min}$ ($M_{\oplus}$) & Plot                                        &  [Fe/H]$_{hot}$ (dex) &  [Fe/H]$_{exo}$ (dex)    \\ \hline
Fiducial    &  5                                             &  9      &  3      &   4.6                                       & Left in Figure \ref{fig3}           &  -0.1                           &   -0.2     \\
Fiducial    &  10                                           &  9      &  3      &   4.6                                       & Right in Figure \ref{fig3}         &  -0.5                           &    \\
Run A       &  3                                             &  8      &  2.5   &   2.5                                       & Upper left in Figure \ref{fig4}  &  0.1                            &  -0.2   \\
Run B      &  10                                           &  9      &  2      &   10                                        & Upper right in Figure \ref{fig4} &  -0.2                          &  -0.2   \\
Run C      &  10                                           &  10    &  3      &   10                                        & Lower left in Figure \ref{fig4}   & -0.4                           & -0.4   \\
Run D      &  10                                           &  10    &  3.5   &   7.2                                       & Lower right in Figure \ref{fig4} & -0.4                           & -0.6    \\
\hline
\end{tabular}
\end{center}
\end{minipage}
\end{table*}

\subsection{The mass of cores for Jovian planets} \label{resu_2}
 
Figure \ref{fig3} (middle) shows the values of the critical core masses $\braket{M_{c,crit}}$ for hot and exo-Jupiters as a function of metallicity 
(see the dashed and solid lines, respectively). 
For comparison, $\braket{M_{c,crit}}$ for low-mass planets is also shown in the same panel (see the dotted line). 
The results show that $\braket{M_{c,crit}}$ for both kinds of Jovian planets is an increasing function of metallicity for both cases of $M_{c,crit0}$ 
(see the left and right panels).   
Since the timescale of gas accretion onto planetary cores shortens for massive cores, 
they have more time to grow up to gas giants. 
This is the main reason why massive planets are preferentially formed for high metallicity stars. 

The results in this panel also show that the value of $\braket{M_{c,crit}}$ of the exo-Jupiters is larger than that of the hot Jupiters. 
(see the middle in Figure \ref{fig3}).  
In our planet trap picture, the mass of the cores tends to increase with increasing the distance from the central star. 
This is achieved because planet traps are assumed to be effective until growing planetary cores open up a gap in their disks. 
The gap-opening mass is an increasing function of distance from the central star. 
It is therefore expected that $\braket{M_{c,crit}}$ of the hot Jupiters is smaller than that of the exo-Jupiters. 
In other words, the cores of the exo-Jupiters drop-out from their planet traps earlier because of their more massive cores, 
and hence they have more time to grow up to fully formed gas giants, compared with those of the hot Jupiters. 
Our results in fact support this trend.\footnote{Intriguingly, \citet{bbl14} have recently inferred using the {\it Kepler} data that 
the core mass of planets is likely to be an increasing function of the distance from the central stars. 
Although their data is very few and confined well within the orbital radius of 1 AU, their finding matches the trend of our results.} 

Another very interesting trend in these critical core mass curves is that the low mass curve exceeds both Jovian planet curves at the lowest metallicities.  
As the metallicity increases, the rising exo-Jupiter curve intersects the low mass curve and then flattens out beyond.
At somewhat higher metallicity, the rising hot Jupiter curve then also intersects the low mass planet curve.  
These intersection points are different for the two cases (see the left, middle panel in Figure \ref{fig3}). 
We define the corresponding values of the metallicity at these intersections as [Fe/H]$_{exo}$ and [Fe/H]$_{hot}$, respectively 
- which we henceforth call {\it transition metallicities}, or TMs. 
In Figure \ref{fig3}, these TMs are denoted by vertical solid and dashed lines respectively. 
Note that these intersections in the $\braket{M_{c,crit}}$ domain are reflected by features in the upper panels 
where the rising PFFs cease their rapid increase at the corresponding TMs and begin to flatten out towards higher metallicity. 

The values of the TMs for the different cases are easily read off the figure.  
For the case of $M_{c,crit0}=5M_{\oplus}$ (see the left, middle plane), 
$\braket{M_{c,crit}}$ of the hot Jupiters intersects with that of low-mass planets at [Fe/H]$_{hot}$ $\simeq -0.1$ (the vertical dashed line) and 
$\braket{M_{c,crit}}$ of the exo-Jupiters does at [Fe/H]$_{exo}$ $\simeq -0.2$ (the vertical solid line).  
Note that in the observed range of metallicities we are modelling, there are two intersections for the case of $M_{c,crit0}=5M_{\oplus}$,  (see the left, top panel). 
For the case of $M_{c,crit0}=10 M_{\oplus}$ (right, middle panel), 
there is only one intersection at [Fe/H]$_{hot}$$\simeq -0.5$ for the hot Jupiters (the vertical dashed line). 

\subsection{Transition metallicities (TMs) and two important masses} \label{two_masses}

Our results indicate that there are two important masses of planets 
for estimating the number of TMs that occur somewhere in the range [Fe/H]$\simeq -0.2$ to -0.4 in $\braket{M_{c,crit}}$-metallicity diagram. 
One of them is obviously the $M_{c,crit0}$ parameter that regulates the onset of gas accretion onto planetary cores (see equation (\ref{m_ccrit})). 
The other is the minimum mass of planetary cores ($M_{c,min}$) that allows the cores to accrete enough amount of gas within a certain time $\tau_{fin}$, 
so that they eventually become gas giants (see equation (\ref{m_cmin})). 

For the case  $M_{c,crit0}=10M_{\oplus}$, one finds that  $M_{c,min} < M_{c,crit0}$ (see Table \ref{table4}).
In this regime, gas giant formation proceeds efficiently even in the low metallicity environment. 
More physically, the timescale of the oligarch growth becomes longer for disks with lower metallicities. 
Nonetheless, the final mass of planetary cores can reach $\ga M_{c,min}$. 
Therefore, gas accretion that takes place after the core formation could be done within the disk lifetime. 
This is because the timescale of gas accretion depends only on the mass of cores (see equations (\ref{tau_KH}) and (\ref{gas_acc})). 
In other words, the total timescale in which the formation of gas giants is complete can still be short, compared with the disk lifetime. 
Thus, $\braket{M_{c,crit}}$ of both the hot and exo-Jupiters becomes larger than that of low mass planets even for disks with low metallicities, 
and hence there is only a TM at low values of [Fe/H] (see the right, middle panel).

For the case of $M_{c,crit0}=5M_{\oplus}$, we obtain $M_{c,min} \simeq M_{c,crit0}$. 
Although we can apply the above argument for this case as well, 
it is expected that the formation of gas giants is not so efficient for disks with low metallicities. 
This occurs because $M_{c,crit0}=5M_{\oplus}(\simeq M_{c,min})$, 
so that the final mass of cores formed in low metallicity disks would be smaller than $M_{c,min}$. 
As a result, such cores cannot grow to gas giants.
Thus, TMs for this case appear in higher metallicity environments than the case of $M_{c,crit0}=10M_{\oplus}$, 
and two TMs exist (see the left, middle panel).

\subsection{PFFs for low mass planets}

We now turn our attention to the metallicity dependence of the PFFs for the low mass planets - 
as well as the total PFF which is the sum of all three populations. 
These plots are given in the bottom panels in Figure \ref{fig3}.   
It is evident that the values of the PFFs for the low mass planets are rather high and quite insensitive to metallicity. 
This indicates that a large number of low-mass planets can be formed by the same mechanism that builds the cores
of the gas giants, over the entire observed range of metallicity. 
This generalizes our conclusion in HP13 that such low mass planets can be regarded as failed Jovian cores and/or mini-gas giants.

On closer examination, another feature of the PFFs of low-mass planets is 
that they descend towards a minimum and then rise again with increasing metallicity.   
This dip has a counterpart in the behaviour of the PFFs of both the hot and exo-Jupiters  (see the top panels), 
which undergo a strong rise in the PFF in this range of metallicities.  
Note that the total PFF for the sum of all populations shows a steady rise without any feature.
Clearly  the planetary cores that might have ended up in low mass planets seed the  growth of gas giants efficiently around TMs. 
For higher metallicity disks, both massive and low-mass planets can form simultaneously, 
so that the PFFs of low mass planets get back to larger values. 

Although similar behaviour is observed for different values of $M_{c,crit0}$ (see the left and right bottom panels), 
there are some quantitative differences. 
The dip in the PFFs is more prominent for the case of $M_{c,crit0}=10 M_{\oplus}$ than the case of $M_{c,crit0}=5M_{\oplus}$. 
This can be understood by the same argument developed in Section \ref{two_masses}, namely, 
that when $M_{c,crit0}=10 M_{\oplus}$, 
the final mass of planetary cores becomes massive enough to speed up the subsequent gas accretion process. 
As a result, gas giants can form efficiently even in relatively low metallicity disks. 
We find that the metallicity dependence of the total PFFs is  insensitive to the value of $M_{c,crit0}$ (see the thick line in the bottom two panels). 
Therefore, such efficient formation of gas giants for the case of $M_{c,crit0}=10 M_{\oplus}$ lowers the PFFs of low mass planets more than 
the case of $M_{c,crit0}=5M_{\oplus}$.

Note that the overall value of the PFFs for the low-mass planets is comparable to or slightly lower than those for the Jovian planets. 
The recent observations, on the contrary, infer that the low-mass planets are the most dominant populations.  
Given that both {\it Kepler} and radial velocity surveys are weighted towards sunlike stars, 
which makes our model applicable to these observations, 
one explanation might be that there are a number of mechanisms for forming the low-mass planets (see Section \ref{disc3}).
However, we note that this may still be a consequence of the effect of stellar masses on the statistics.
We need to perform similar calculations along the initial mass function (IMF) 
and to examine how different the resultant PFFs are for different stellar types. 
This is because low-mass stars, that tend to generate more low-mass planets due to low-mass disks, 
occupy the dominant contribution in the standard IMF.
This kind of approach may be useful for microlensing surveys which tend to focus on low-mass stars.

The important conclusion from all of this is that the dominant planetary populations switch from low-mass planets to massive ones at the TMs (see Figure \ref{fig3}). 

\section{Parameter study} \label{para}

As discussed above, our results show that 
the PFFs (and TMs) of both the hot and exo-Jupiters around [Fe/H]$\simeq -0.2$ to -0.4 behave differently for different values of $M_{c,crit0}$ 
with $c$ and $d$ fixed (see Figure \ref{fig3}). 
This suggests that we can derive important constraints on the value of $M_{c,crit0}$ by exploring different values of $c$ and $d$. 
As pointed above, these parameters are likely to have a range of values: $8 \la c \la 10$ and $2 \la d \la 4$.

\subsection{Robustness of the planet-metallicity relation} 

We now examine how our PFFs and critical core masses are altered by changing the values of $M_{c,crit0}$, $c$ and $d$, 
which control the efficiency of planet formation (see equations (\ref{m_ccrit}) and (\ref{tau_KH})). 
The different model parameters are given in  Table \ref{table4}. 
We choose the values of $M_{c,crit0}$, $c$ and $d$, so that $M_{c,crit0}$ satisfies the condition $M_{c,min} \simeq M_{c,crit0}$ (see Table \ref{table4}). 
As discussed in Section \ref{two_masses} and will be shown below, 
TMs appear in the range [Fe/H]$\simeq$ -0.2 to -0.4 when this condition is met.
Note that this was confirmed by performing more than 30 runs.

Figure \ref{fig4} shows some of our results. 
The upper, left panel shows the results for the case $M_{c,crit0}=3M_{\oplus}$, $c=8$, and $d=2.5$ (Run A), 
the upper, right panel is for $M_{c,crit0}=10M_{\oplus}$, $c=9$, and $d=2$ (Run B), 
the lower, left panel is for $M_{c,crit0}=10M_{\oplus}$, $c=10$, and $d=3$ (Run C), 
and the lower, right panel is for $M_{c,crit0}=10M_{\oplus}$, $c=10$, and $d=3.5$ (Run D).  

The resultant PFFs show the same trends for a wide range of $M_{c,crit0}$, $c$ and $d$ (see the top figure in each panel), 
as  those observed for the case of $M_{c,crit0}=5M_{\oplus}$, $c=9$, and $d=3$ shown in Figure 3.  
The shape of the PFFs for the two Jovian class planets has the same general structure.  
The exo-Jupiters always dominate the hot Jupiter population  and after a rather steep rise from lower metallicity,
both PFFs roll over into very gently increasing values beyond the exo-TM.  
The PFFs for the low mass planets all have dips and drop significantly around [Fe/H]$\simeq-0.2$ to -0.4, 
and rise again to near their original values at higher metallicity.   
This is the hallmark of the planet-metallicity relation.   
The other general result is that at low metallicity, the low mass planets dominate the total population for all models.  
It is impressive how robust these relations are over such a wide range of the basic parameters of the model.  

\subsection{Trends in the TMs} 

The obvious difference between the models in our parameter study is the behaviour of the TMs. 
For Run A, two TMs appear around [Fe/H]$_{hot}\simeq0.1$ and [Fe/H]$_{exo} \simeq-0.2$ for the hot and exo-Jupiters, respectively. 
For Run B, there is only one TM at  [Fe/H]$_{hot,exo} \simeq-0.2$. 
For Run C, only one TM exists around [Fe/H]$_{hot,exo} \simeq-0.4$. 
For Run D, there are two TMs located around [Fe/H]$_{hot} \simeq-0.4$ and [Fe/H]$_{exo} \simeq -0.6$ for the hot and exo-Jupiters, respectively.     
The TMs focus on properties of the cores which give us important additional information than the behaviour of the PFFs.  
Why do the positions (in metallicity) and numbers  of TMs  change? 

\begin{figure*}
\begin{minipage}{17cm}
\begin{center}
\includegraphics[width=7.5cm]{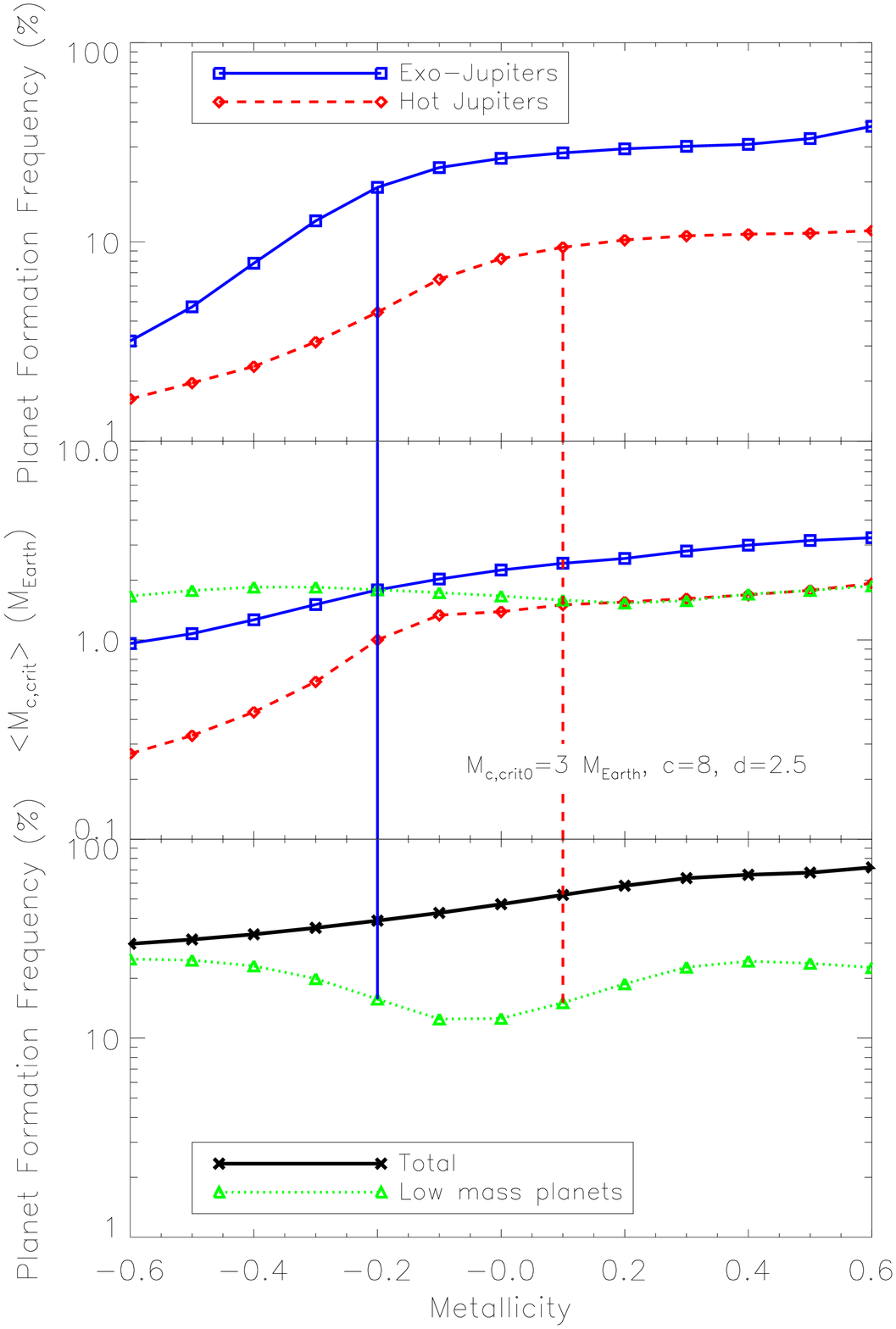}
\includegraphics[width=7.5cm]{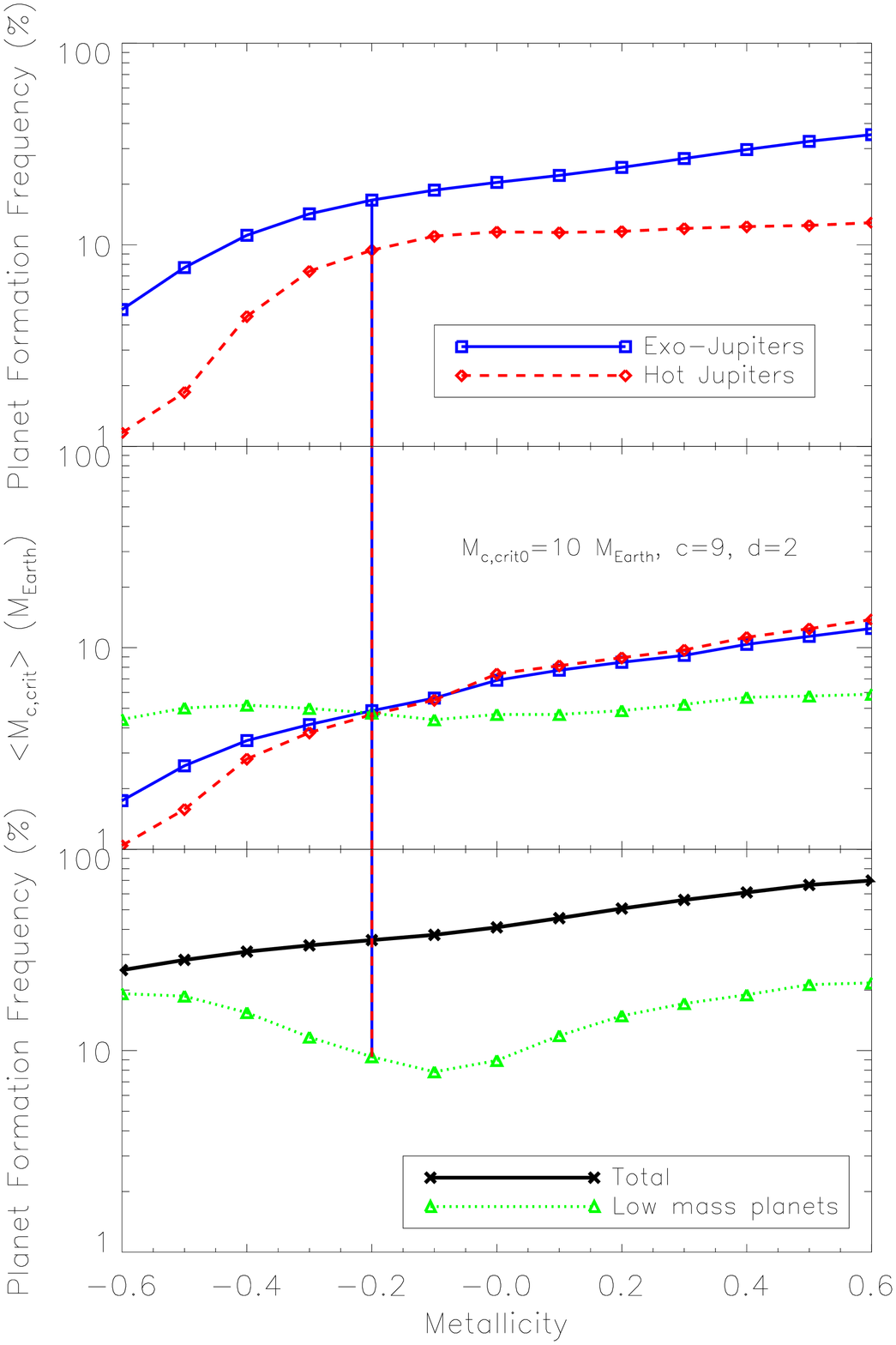}
\includegraphics[width=7.5cm]{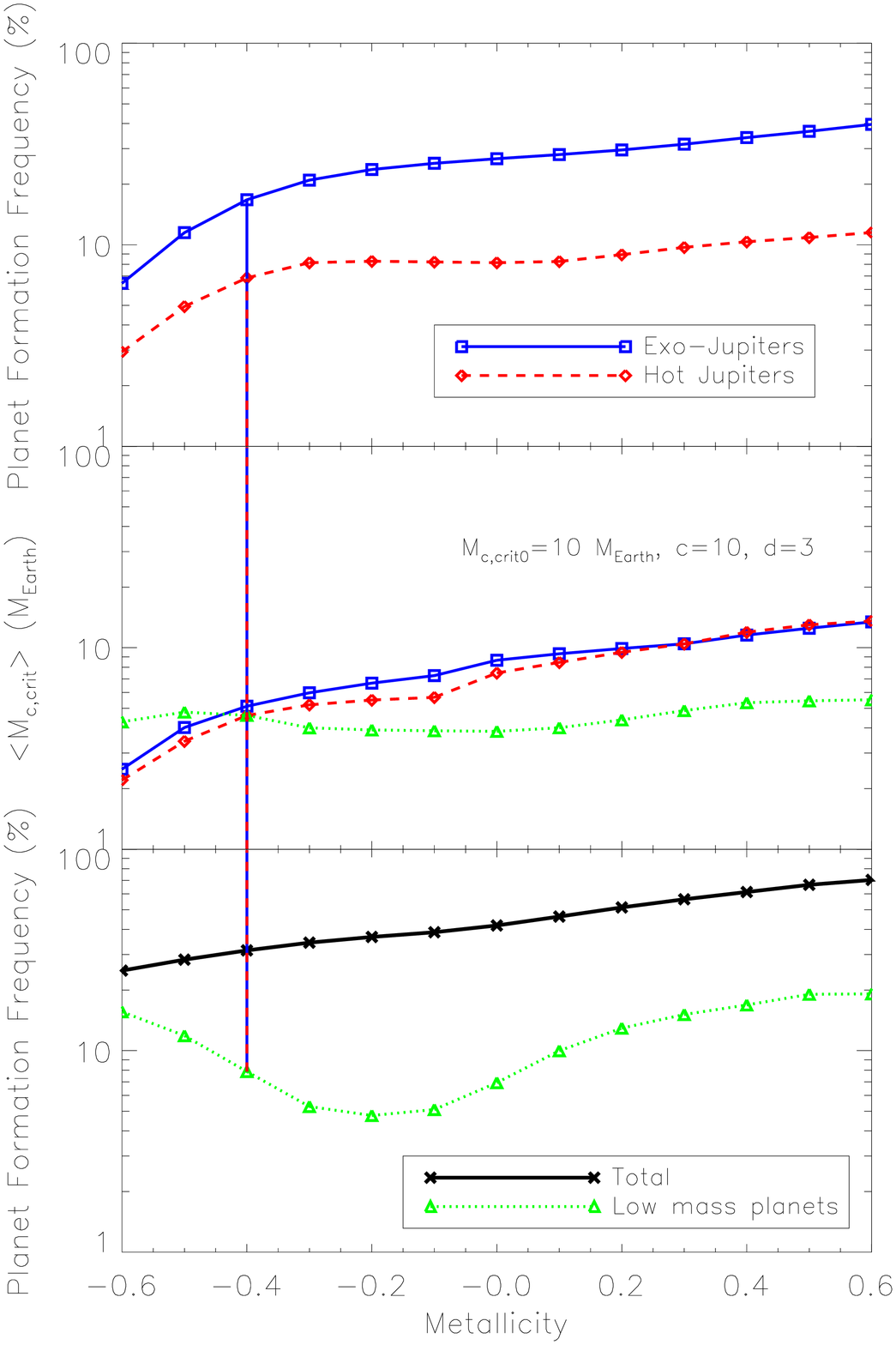}
\includegraphics[width=7.5cm]{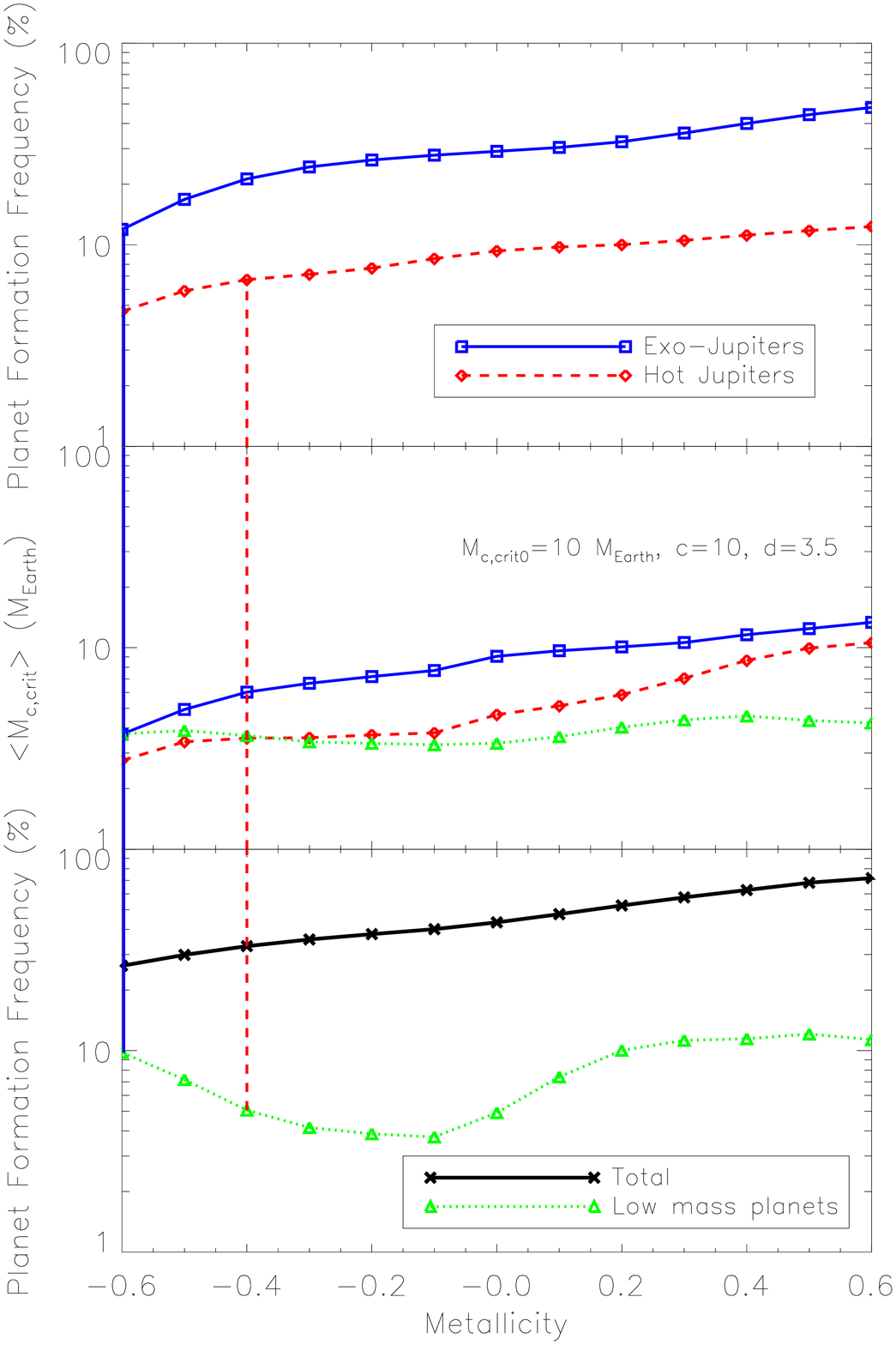}
\caption{The resultant PFFs and the mean mass of planetary cores as a function of the metallicity (as Figure \ref{fig3}, also see Table \ref{table4}). 
The results for Run A ($M_{c,crit0}=3M_{\oplus}$, $c=8$, and $d=2.5$) are shown in the upper, left panel, 
those of Run B ($M_{c,crit0}=10M_{\oplus}$, $c=9$, and $d=2$) are on the upper, right panel, 
those of Run C ($M_{c,crit0}=10M_{\oplus}$, $c=10$, and $d=3$) are on the lower, left panel, 
and those of Run D ($M_{c,crit0}=10M_{\oplus}$, $c=10$, and $d=3.5$) are on the lower, right panel. 
Although all these results of the PFFs are very similar to the case of $M_{c,crit0}=5M_{c,crit0}$, $c=9$, and $d=3$, 
the degeneracies can be resolved if the position and the number of TMs are examined in detail (see the vertical lines).}
\label{fig4}
\end{center}
\end{minipage}
\end{figure*}

First consider the positions.  
These are affected by the sensitivity of $\braket{M_{c,crit}}$ for both Jovian planets, to the value of $M_{c,crit0}$.  
As shown in Figures \ref{fig3} and \ref{fig4}, 
the magnitude of $\braket{M_{c,crit}}$ for both Jovian planets is almost linearly proportional to the value of $M_{c,crit0}$. 
As an example,  the value of $\braket{M_{c,crit}}$ for the exo-Jupiters at  solar metallicity is approximately equal to the value of $M_{c,crit0}$.  
On the contrary, the value of $\braket{M_{c,crit}}$ for the low-mass planets is quite insensitive to the value of $M_{c,crit0}$. 
Our results show that $\braket{M_{c,crit}} \simeq 2-4 M_{\oplus}$ for various values of $M_{c,crit0}$.
As a result, the locations of the TMs generally sweep towards lower metallicity with increasing $M_{c,crit0}$.

What about the difference in the number of the TMs?   
There is only one TM for Runs B and C. 
Note that for these two cases, the curves for $\braket{M_{c,crit}}$ as a function of metallicity of both the hot and exo-Jupiters 
almost completely line up with one another.  
In this case, it is obvious that both intersect the low mass curve at one metallicity - hence there is one TM.   
Why do the curves  line up in these cases?
Based on a large number of similar simulations (which are not shown here), 
we found that this is a consequence of a large value of $M_{c,crit0}(=10M_{\oplus})$. 
As discussed below, this match is attributed to the cloud coupling of planetary migration with planetary growth.

As discussed in Section \ref{resu_2}, $\braket{M_{c,crit}}$ of the hot Jupiters is generally smaller than that of the exo-Jupiters. 
Most of our results obviously support the trend (see the middle panel in Figures \ref{fig3} and \ref{fig4}). 
For the case of $M_{c,crit0}=10M_{\oplus}$, however, 
the critical mass curves for the hot Jupiters are raised up to the values of the exo-Jupiter curves for Runs B and C,
(but not the two other cases). 
In this more massive regime, the final mass of planetary cores can become larger than the gap-opening mass. 
This leads to switch of core migration from trapped type I to type II, even as  the formation of the planetary cores still proceeds.  
As a result, the distribution of cores that is generated by planet traps can be washed out, 
making it easier for the core mass curves of these populations to line up with one another.  

It is interesting to investigate further the coupling of migration with planet formation, especially focusing on type II migration as well as gas accretion onto cores.
When the results of Run C are compared with those of Run D, there is the difference in the behavior of $\braket{M_{c,crit}}$ for both Jovian planets, 
although the value of $M_{c,crit}$ itself is the same for these two cases.
For the case of Run C, $\braket{M_{c,crit}}$ for both Jovian planets lines up entirely (only one TM exists) 
whereas for Run D, the value is different for different populations (two TMs exist). 

The difference can be understood by the combined effects of the gas accretion and type II migration. 
If the gas accretion proceeds quickly (equivalently for a small value of $M_{c,min}$, e.g., see Run D), gas giant formation is complete earlier. 
Then the inertia of such fully formed planets becomes important for slowing down type II migration significantly. 
This eventually minimizes the effects of type II migration, 
and hence the final distribution of massive planets is affected largely by the distribution of planetary cores generated by planet traps. 
Consequently, $\braket{M_{c,crit}}$ of the hot Jupiters tends to be smaller than that of the exo-Jupiters (see the lower right panel in Figure \ref{fig4}). 
When planetary cores accrete gas slowly (equivalently for a large value of $M_{c,min}$, e.g., see Run C), 
they need more time to form gas giants. 
This, in turn, maximizes the importance of type II migration. 
Eventually, the effects of planet traps are washed out and $\braket{M_{c,crit}}$ of the hot Jupiters becomes comparable to that of the exo-Jupiters 
(see the lower left panel in Figure \ref{fig4}).

Thus, similar results in the PFFs can be obtained for a wide range of $M_{c,crit0}$, $c$, and $d$. 
Nonetheless, such degeneracies may be able to be resolved 
if the position and the number of TMs in the $\braket{M_{c,crit}}$-metallicity diagram are examined carefully. 
This occurs because planet formation is intimately coupled with planetary migration. 

\subsection{The best fit value for $M_{c,crit0}$ } \label{comp}

Having demonstrated the planet-metallicity relation is well described by the trend in the PFFs,
what value for $M_{c,crit0}$ is most compatible with the data?   It turns out that the distributions of hot and 
exo-Jupiters as a function of metallicity can provide an important and rather robust constraint, as we now show. 

\begin{figure*}
\begin{minipage}{17cm}
\begin{center}
\includegraphics[width=8cm]{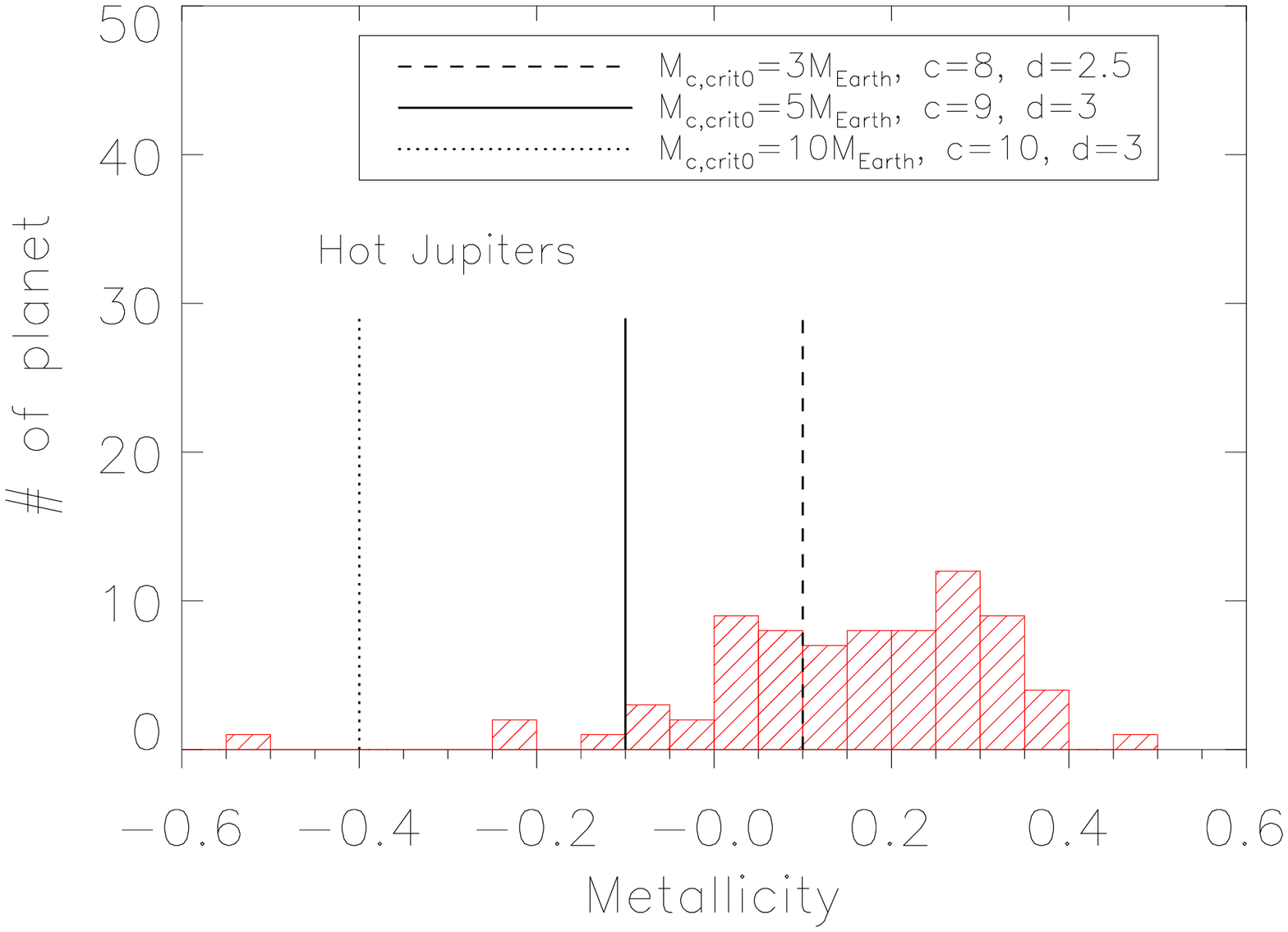}
\includegraphics[width=8cm]{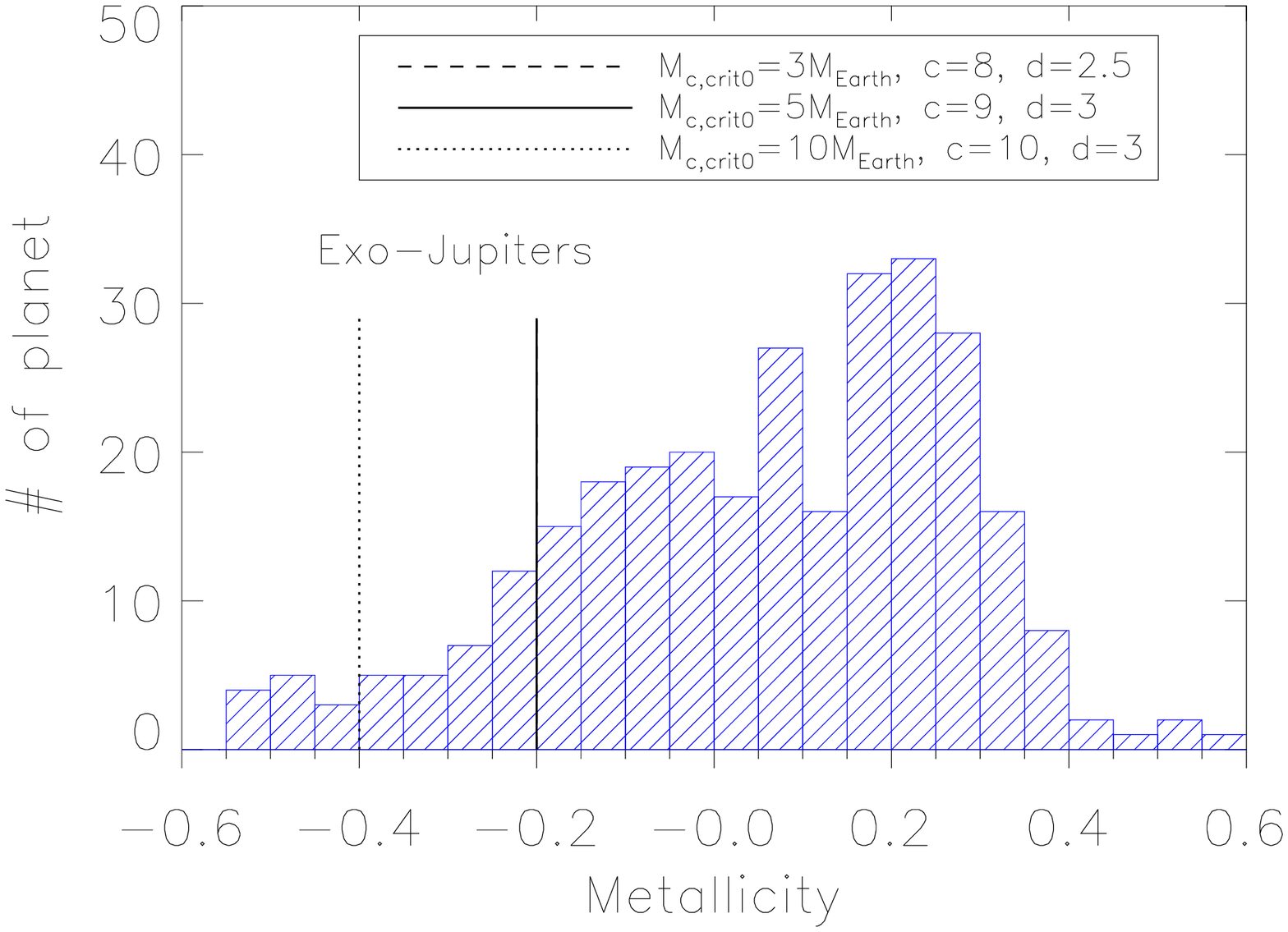}
\caption{The comparison of our results with the radial velocity observations. 
The number of the hot and exo-Jupiters is replotted in the histogram (also see Figure \ref{fig1}). 
The values of TMs for both the hot and exo-Jupiters are also shown (see the vertical lines, also see Table \ref{table4}). 
The dashed lines denote for the case of $M_{c,crit0}=3M_{\oplus}$, $c=8$, and $d=2.5$, 
the solid ones are for the case of $M_{c,crit0}=5M_{\oplus}$, $c=9$, and $d=3$, 
and the dotted ones are for the case of $M_{c,crit0}=10M_{\oplus}$, $c=10$, and $d=3$. 
Note that the dashed line on the right panel completely match the solid line. 
Our results imply that the currently available data of exoplanets are likely to be fitted well by $M_{c,crit0}$ that is smaller than $10M_{\oplus}$.}
\label{fig5}
\end{center}
\end{minipage}
\end{figure*}

We first discuss the observations of exoplanets. 
In Figure \ref{fig5},  the observational data are replotted 
to show the distributions of hot and exo-Jupiters as a function of metallicity in the left and right panels respectively. 
We here focus in particular on the radial velocity data for the following reasons (also see Section \ref{disc5}). 
First, exoplanets at $r > 1$ AU are currently detected mainly by the radial velocity methods. 
Thus, picking up only the radial velocity data enables us to use the observational data that are consistent at least on the detection method  
for both the hot and exo-Jupiters. 
Second, the resultant populations are in good agreement with the current knowledge of the statistical properties of gas giants: 
the hot Jupiters are rare and the exo-Jupiters are dominant in the population of massive planets. 
Figure \ref{fig5} infers that the hot Jupiters may be more sensitive to metallicity than the exo-Jupiters.  
While it seems that the observational data are still not large enough to examine this trend statistically, 
some clues about it have recently been reported \citep{bn13,afs13}.

Figure \ref{fig5} also shows, by means of vertical lines, the TMs  for both the hot and exo-Jupiters 
([Fe/H]$_{hot}$ and [Fe/H]$_{exo}$) above which the PFFs for them will increase significantly (see values in Table \ref{table4}). 
In particular, we select the cases of $M_{c,crit0}=3M_{\oplus}$, $c=8$, and $d=2.5$ (see the dashed line), of $M_{c,crit0}=5M_{\oplus}$, $c=9$ and $d=3$ 
(see the solid line), and of $M_{c,crit0}=10M_{\oplus}$, $c=10$, and $d=3$ (see the dotted line). We note that the dashed line for the exo-Jupiters lines-up 
with the solid one completely (see the right panel).

Since the TM values mark the metallicities above which the populations of hot and exo-Jupiters begin to dominate the low
mass planets, it is clearly evident that the choice of the value $M_{c,crit0}=10M_{\oplus}$ for the hot Jupiters puts the
TM far to the 
left of the distribution.    For the exo-Jupiter population, this value puts the TM in the midst of the distribution.
On the other hand, the choice of  a value $M_{c,crit0}=3M_{\oplus}$ puts the TM for hot Jupiters in the midst of the population,
with a significant amount of power at the lower metallicities.  This is also problematic.   It is clear that the optimal choice
for this parameter is $M_{c,crit0}=5M_{\oplus}$.   For the hot Jupiters, this places the TM just to the left of the rise in the 
population of observed population of hot Jupiters, and at a reasonable TM for the exo-Jupiters as well.
Our results therefore indicate that the case, $M_{c,crit0}=5M_{\oplus}$, $c=9$ and $d=3$ is the most consistent match with the currently 
observed metallicity distributions derived for hot and exo-Jovian planets detected by radial velocity observations.

It is particularly interesting that the observations pick out the value; $M_{c,crit0}=5M_{\oplus}$.   
This best fit model is the fiducial case shown on the left panel of Figure \ref{fig3}. 
If we examine the PFF for the exo-Jupiter curve in the top panel of this Figure, we note that at solar metallicity, the most probably core mass
of such a planet is, 
\begin{equation}
\braket{M_{c,crit}} \simeq 5 M_{\oplus}
\end{equation}
which is the value of our parameter, $M_{c,crit0}$.   Therefore our best fit critical core mass parameter - given the available data on the 
distributions of metallicities of Jovian planets - is indeed behaving like
a physical critical core mass in our model.  This shows that our model is internally self-consistent.  

\section{Discussion}  \label{disc}

Our parameter study has a number of implications for observed solar and extrasolar planets including our own Jupiter as well as super-Earth planets. 
Two of our general conclusions; that the dense population of Jovian planets at 1AU is a consequence of planet trap dynamics and not metallicity;
and that the critical core mass for runaway gas accretion is fundamental for our models, may be both open to future direct observational tests.
In addition it is interesting to reconsider formation mechanism of hot Jupiters, based on our calculations.
We discuss these implications of our models, as well as caveats.

\subsection{The mass of Jupiter's core} \label{disc1}

If our own Jupiter  can be regarded as a typical member of the exo-Jupiter population, then our model
predicts that it likely has a core mass of the order of 5 $M_{\oplus}$.   Can this be justified?

On the theoretical side, we recall that $M_{c,crit0}$ is directly related to the opacity in the planetary envelope \citep[equation (\ref{m_ccrit0}), also see][]{hi10}. 
A value of $5M_{\oplus}$ implies that the grain opacity in the atmosphere of forming Jupiters must have a characteristic value 
that is more than one order of magnitude lower than the canonical value;  $\kappa \simeq 0.06 $ g cm$^{-2}$. 
Recently, \citet{mp08} have shown through numerical simulations 
that the resultant opacity in planetary atmospheres is very unlikely to depend strongly on the metallicity of accreting materials. 
This suggests that the growth and subsequent settling of dust grains in the envelope play an important role in lowering the grain opacity there. 
Such effects are in fact confirmed by \citet{mbp10}, 
wherein more complete calculations of planetary atmospheres coupled with grain growth and settling there were undertaken.
This detailed study implies that the grain opacity can be reduced to $\leq $2 \% of the conventional value 
(equivalently a factor of $\geq$ 50). 
Note that the difference in the value of $\kappa$ may not be crucial because the dependence of $\kappa$ in equation (\ref{m_ccrit0}) is very shallow. 
In fact, the recent population synthesis calculations suggest that, 
while it is very likely that the grain opacity in the planetary atmosphere is much lower than the ISM value, 
it is very difficult to constrain $\kappa$ sharply due to such dependence \citep{mka14}.

In addition, our results show that 
the PFFs of low mass planets for the case of $M_{c,crit0}=5M_{c,crit0}$ are less sensitive to those for $M_{c,crit0}=10M_{c,crit0}$ 
(see Figures \ref{fig3} and \ref{fig4}). 
It is interesting that $\braket{M_{c,crit}}$ of low mass planets for both cases of $M_{c,crit0}$ is very similar, 
which is about $2-4M_{\oplus}$ (see Figures \ref{fig3} and \ref{fig4}). 
This indicates that it may be difficult to derive invaluable constraints on the core mass of Jovian planets by examining the mass of super-Earths and hot Neptunes. 

Detailed numerical simulations of planetary envelopes show 
that the accretion of solid materials can proceed efficiently in the final, runaway gas accretion phase \citep{p96,hbl05,mbp10}. 
Based on the orbital integration of planetesimals around a planet done by \citet{si08}, 
the additional accumulation of the solids can become about serval $M_{\oplus}$. 
Thus, a 5 $M_{\oplus}$ core mass for the Jupiter estimated by our model is obviously the lower limit.

\subsection{Origins of hot Jupiters}

As discussed above, the hot Jupiters are produced as a consequence of the orbital migration induced by planet-disk interactions in our model. 
In fact, the first discovery of the massive planet around the Sun-like star \citep{mq95} invokes the importance of gas-induced planetary migration \citep{lbr96}. 
On the other hand, the recent observations of the Rossiter-McLaughlin effect have detected close-in massive planets 
whose orbital planes are strongly misaligned with those of their host stars \citep[e.g.,][]{tcq10,wfa10}. 
It is suggested that such misalignment can be understood by planet-planet interactions with the aid of convergent gas-induced migration 
and/or the Kozai mechanisms combined with stellar tides \citep[e.g.,][]{ft07,nib08,cfm08}.
In addition, the recent exoplanet observations infer that 
the formation of the hot Jupiters in higher metallicity disks may be more efficient than our results predict \citep[e.g.,][]{sim00,fv05,dm13}.
Furthermore, the hot Jupiters are subject to the exposure of the strong irradiation of the host stars 
which can considerably affect the composition and the structure of the atmosphere of the planets \citep[e.g.,][]{sd10}. 
In order to fully understand origins of the hot Jupiters, 
it is therefore important to consider additional physical processes that will occur after the dissipation of gaseous disks 
in which gas giant formation (including the hot Jupiters) predominately proceeds.

\subsection{Implications for low-mass planets in tight orbits - super-Earths} \label{disc3}
 
How do low-mass planets  form? 
Our results on the critical core mass of massive planets of $5 M_{\oplus}$ have important implications 
for low-mass planets in tight orbits such as super-Earths.

Currently, there are probably two main ideas to understand the formation of low-mass planets: one is that they form in-situ, 
the other is that they are formed in the outer part of disks and migrate to the current position. 
Since the former scenario considers mergers of embryos and planetesimals as the main assembly process to form super-Earths \citep[e.g.,][]{hm13,cl13}, 
the finally formed planets are predicted to be scaled-up version of rocky planets like our Earth. 
The latter picture emphasizes that such low-mass planets can form by the same mechanism as gas giants, 
and that therefore they are essentially failed cores of gas giants and/or mini-gas giants \citep[e.g.,][HP12, HP13]{rbl11}. 
Note that the in-situ scenario requires unusually large amount of solids in the inner part of disks, 
so that it is very likely that some kind of migration would be needed 
to transport embryos of planets from the outer to the inner part of disks \citep[e.g.,][]{oi09,mn10,kl12}.

The statistical investigation presented in this paper can provide additional insight into the failed core scenario.
We predict that any core mass beyond $5 M_{\oplus}$ will have an increasing gaseous component, 
since gas accretion takes place efficiently when $M_c > M_{c,crit} (\simeq 5M_{\oplus})$. 
As a result, we can expect that super-Earths beyond $5 M_{\oplus}$ will not be composed of pure solid cores, 
but will have a greater admixture of gas as one goes from $5-10 M_{\oplus}$.  
This may give a clue to understand the recent results of the {\it Kepler} observations 
which derive a new empirical relation between planetary mass and radius that  
infers that the composition of super-Earths may change from solid materials to gaseous ones around the radius of planets, 
$R \simeq 1.5 R_{\oplus}$ (equivalently the mass of planets, $M_p \simeq 5 M_{\oplus}$ using the relation) \citep{wm14,mih14}.   

Note that low-mass planets in tight orbits are anticipated to be influenced by stellar irradiation like hot Jupiters, 
which affects the abundance and the volume of the atmosphere \citep{rbl11,lfm12,lf13}. 
Thus, it would also be important to consider the evolution of such planets to interprets their current properties more accurately.

\subsection{Hierarchy of planetary types}

Combining our results with the above discussion, we suggest the following hierarchy for planets formed in protoplanetary disks.

\begin{itemize}
\item Massive planets ($M_p\geq 10 M_{\oplus}$): cores ($\simeq (f + 5) M_{\oplus}$) + massive envelopes ($\geq 5 M_{\oplus}$).
\item Intermediate-mass planets ($5M_{\oplus} \leq M_p\leq 10 M_{\oplus}$): cores ($\simeq 5 M_{\oplus}$) + low-mass envelopes ($< 5 M_{\oplus}$).
\item Low-mass planets: only cores ($M_p \leq 5 M_{\oplus}$).
\end{itemize}

Note that $f$ in the core mass of massive planets arises from the additional capture of solid materials during the runaway gas accretion (see Section \ref{disc1}).

In the hierarchy, the difference between gas giants like the Jupiter and sub-giants like the Neptune originates from the timing of disk dissipation.
It determines the timescale during which growing planets can accrete the disk gas. 
Thus, when planets are formed in the later phase of disk evolution, they tend to be sub-giants.
The regime of super-Earths covers both the intermediate and the low-mass planets. 
This implies that they can be regard as both failed cores of gas giants as well as massive Earths. 
It is therefore expected that the composition of them would be diverse.
Interestingly, \citet{bbl14} have recently examined the {\it Kepler} data as a function of metallicity, 
and inferred that there are likely to be three types of exoplanets with different compositions. 
Although more observational data would be needed, 
our proposed hierarchy roughly corresponds to their classification, 
by converting the planetary size to the mass using an empirical mass-radius relation.
More detailed studies are obviously required for examining the proposed hierarchy.

\subsection{Caveats} \label{disc5}

There are two caveats for our results.  
The first is related to the state of the observational data sets that we have employed.
The currently available observations of exoplanets, especially around low metallicity stars, 
may still be highly biased even for the most successful radial velocity methods.   
Thus, it is very important to conduct intensive surveys of exoplanets, especially orbiting around stars with [Fe/H]$\simeq-0.2$ to -0.4, 
to create a more complete sample of the distribution of massive planets at these values of [Fe/H]. 
This would be invaluable information for our model, 
since it would allow us to better constrain the parameter values, using the approach that we have outlined here.   
As a concrete example, if more data were to fill out the metallicity distribution in Figure \ref{fig5} for the hot Jupiters towards lower values of the metallicity,
this would tend to push the value of $M_{c,crit0}$ towards a higher values like $10 M_{\oplus}$.   
In addition, more data for metal-poor stars may clarify the difference in the distribution of massive planets: 
the observed hot Jupiters show a sharp transition in Figure \ref{fig5} 
whereas, for the exo-Jupiters, neither the computed PFFs nor the observed ones show such a sharp trend.

The second has to do with our theoretical evolutionary tracks.  
We have adopted an approximation, in common with the Ida \& Lin models 
that gas accretion onto critical cores occurs at a rate that is governed by the Kelvin-Helmholtz timesscale (see equations (\ref{tau_KH}) and (\ref{gas_acc})).  
We find that the resulting masses of our Jovian planets are underestimated by this approach.  
We hasten to note that this is not likely to affect any of the conclusions related to the PFFs as is shown by figure 4 in HP13.  
There we found that even if the amount of accreted gas onto planetary cores changes drastically,
the high mass regime did not significantly affect the PFFs.   
There is some sensitivity however to the low mass planets, whose PFFs are an overestimate if accretion rates are increased.  
We note that other authors have adopted what is surely an extreme upper limit to the planet accretion rate - 
namely the full disk accretion rate \citep[e.g.,][]{mab09}.  
In that limit, there is of course no problem to form very massiveJupiters, although the low mass planets are entirely missed.  
More theoretical work is needed to formulate a more accurate description of the physics of late-phase gas accretion \citep[also see][for a recent discussion]{hi13}.  

\section{Conclusions} \label{conc}

We have investigated quantitatively the sensitivity of the core accretion scenario to metallicity 
with the goal of understanding the planet-metallicity correlation. 
To this end, we have adopted a semi-analytical model 
that was developed in a series of our papers (HP11, HP12, HP13), 
wherein evolutionary tracks of planets growing in planet traps are constructed. 
We have considered three types of planet traps: dead zones, ice lines and heat transitions (HP11). 
As in HP13, we have utilized the tracks for estimating PFFs that are calibrated in specific zones in the mass-semimajor axis diagram. 
An important aspect of building models of Jovian planet formation is the origin of planetary cores and their sensitivity to metallicity, 
and we have constructed a very useful approach for studying this as well.  
In this paper, we have focused on three zones in the mass-semimajor axis diagram,  
in which most dominant exoplanet populations are located (see Table \ref{table1}). 
We have computed the PFFs as well as the behavior of the critical core masses ($M_{c,crit}$) for planets that end up in these three zones, 
as a function of the metallicity. 
In our formalism, the change of metallicity is directly related to the variation of the dust density in disks (see equation (\ref{z_disk})). 
We list our conclusions below. 

\begin{enumerate}

\item The PFFs of the exo-Jupiters are higher than those of the hot Jupiters for a wide range of metallicity 
(see the top panels of Figures \ref{fig3} and \ref{fig4}). 
This is consistent with the findings of HP12. 
Thus, we can conclude that planet traps coupled with core accretion lead to a denser population of gas giants around 1 AU, 
and a smaller population of hot Jupiters, for the entire range of metallicities.  

\item The PFFs of both the hot and exo-Jupiters have low values at low metallicity and rise rather steeply through this regime.  
At transition metallicities (TMs) of the ranges [Fe/H]$\simeq -0.2$ to -0.4,  the PFFs of both the hot and exo-Jupiters start to level off
and are less sensitive to increasing metallicity (see the top panels of Figures \ref{fig3} and \ref{fig4}).  
The PFFs for low mass planets, on the other hand, dominate the population at the lowest metallicities, 
undergo a dip in the trend as metallicity increases through the transition range, 
and then rise up to the initial value at higher metallicity.   

\item The total PFFs undergo a steadily increasing trend, with no strong features, as the metallicity increases.  
These results are in excellent agreement with the general planet-metallicity relation.   
 
\item We have shown that the trend is valid for a wide range of $M_{c,crit0}$ 
that is a parameter in our model involved with the onset of gas accretion (see equation (\ref{m_ccrit0})). 
We have clarified the trend by plotting the mean critical mass of planetary cores ($\braket{M_{c,crit}}$, see equation (\ref{M_c_mean})) 
- the trend is a direct reflection of $\braket{M_{c,crit}}$ which is an increasing function of metallicity. 
As a result, the formation of gas giants proceeds more efficiently for higher metallicity disks.

\item The behaviour of the PFFs of both the Jovian planets is different for different values of $M_{c,crit0}$, 
especially for the low metallicity environment (see Figure \ref{fig3}). 
We have found that for the case of $M_{c,crit0}=5M_{\oplus}$ 
the PFFs of the hot and exo-Jupiters drop rapidly around [Fe/H]$_{hot} \simeq -0.1$ and [Fe/H]$_{exo} \simeq -0.2$, respectively 
whereas those of the hot Jupiters may decrease around [Fe/H]$_{hot} \simeq -0.5$ for the case of $M_{c,crit0}=10M_{\oplus}$. 
We have shown that the rapid decrement is understood by comparing $\braket{M_{c,crit}}$ for massive planets and low mass planets. 
This is because $\braket{M_{c,crit}}$ of low mass planets gives the threshold mass of planetary cores that can eventually grow to gas giants. 
Thus, the comparisons enable one to define TMs below which the PFFs of massive planets decline sharply.

\item We  performed a parameter study in the efficiency of planetary growth. 
More specifically, we have varied the values of $M_{c,crit0}$, $c$, and $d$ (see equations (\ref{m_ccrit}) and (\ref{tau_KH}), also see Table \ref{table4}), 
and examined how our results are altered. 
We have found that the overall features of the PFFs of the both hot and exo-Jupiters are very similar 
if the condition $M_{c,crit0} \simeq M_{c,min}$ is satisfied (see equation (\ref{m_cmin})). 
Nonetheless, TMs behave differently for different sets of $M_{c,crit0}$, $c$, and $d$ (see Figure \ref{fig4}). 

\item The best fit of the parameters to the extant observational data on planetary metallicity distributions is; 
$M_{c,crit0}=5M_{\oplus}$, $c=9$, and $d=3$.  
The degeneracies discussed above may be able to be resolved if the features of TMs are investigated in detail. 
Thus, future observations of exoplanets, especially around [Fe/H]$=-0.2$ to -0.4 are crucial for constraining these three quantities.

\item The PFFs of low-mass planets do not decrease steadily with decreasing metallicities (see the bottom panels of Figures \ref{fig3} and \ref{fig4}). 
In our model, such planets are formed as "failed" cores of gas giants and/or mini-gas giants, 
implying that a large fractions of observed super-Earths and hot Neptunes may be formed by the same mechanism forming massive planets.  
We suggest
on this basis, that SuperEarths exceeding 5 $M_{\oplus}$ will increasingly have gas contributions to their structure.   

\end{enumerate}

Thus, we have shown that the standard core accretion scenario linked with planet traps in disks  shows that SuperEarths and low mass planets 
are the dominant populations at low metallicity that 
provide the foundations for building bigger cores and gas giants as the dust to gas ratio increases.  
In a subsequent paper, we will examine the effects of the stellar mass on the PFFs and planet-metallicity correlation.


\acknowledgments
The authors thank Tristan Guillot, Shigeru Ida, and  Christoph Mordasini for stimulating discussion, 
and a referee, Jack Lissauer, and another anonymous referee for useful comments on our manuscript. 
Also, YH thank the hospitality of Tokyo Institute of Technology for hosting stimulating visits. 
Y.H. is supported by EACOA Fellowship that is supported by East Asia Core Observatories Association which consists of 
the Academia Sinica Institute of Astronomy and Astrophysics, the National Astronomical Observatory of Japan, the National Astronomical 
Observatory of China, and the Korea Astronomy and Space Science Institute.  R.E.P. thanks the Academica Sinica Institute for
Astronomy and Astrophysics (ASIAA) in Taiwan for hospitality during a visit in which the manuscript was 
completed.  His research is supported by a Discovery Grant from the Natural Sciences and Engineering Research Council (NSERC) of Canada.





\appendix

\section{A: Semi-analytical model for the evolution of forming and migrating planets at planet traps} \label{app1}

Here we briefly describe our prescription for forming planets at planet traps.
Table \ref{tableA1} summarizes physical quantities and model parameters involved with our model (also see Table \ref{table2}).
As discussed below, the most uncertain parameter, that may affect our results, is $f_{fin}$ (see Equation (\ref{M_max})).
It is also emphasized that, except for the choice of $f_{fin}$, there is no adjustment for any physical process for generating planetary populations. 

\begin{table*}
\begin{minipage}{17cm}
\begin{center}
\caption{Additional quantities in our model}
\label{tableA1}
\begin{tabular}{ccc}
\hline
Symbol             &  Meaning                                                                                                                                     & Values           \\ \hline
                          &  Physical quantities                                                                                                                     &                                    \\ \hline           
$M_c$               &  The mass of planetary cores computed, following their tracks                                                  &                       \\
$R_c$               &   The radius of planetary cores (see Equation (\ref{r_c}))                                                            &                       \\
$M_{c,sio}$       &  The isolation mass of planetary cores (see Equation (\ref{mc_iso}))                                         &                      \\
$\tau_{c,acc}$   &  The growth timescale for planetary cores (see Equation (\ref{tau_cacc}))                                &                      \\
$\Omega$         &   Orbital frequency                                                                                                                      &                     \\
$\sigma$           &   Velocity dispersion of planetesimals                                                                                          &                     \\
$\triangle r_c$       &   The feeding zone of forming planetary cores ($=b r_H$)                                                          &                     \\
$r_H$                &   The Hill radius of planetary cores ($=(M_c/(3M_*))^{1/3}$)                                                      &                     \\
$H$                   &    Disk pressure scale height                                                                                                       &                     \\
$\nu$                 &   Disk viscosity ($=\alpha H^2 \Omega$)                                                                                   &                 \\
$M_{gap}$         &   The gap-opening mass (see Equation (\ref{M_gap}))                                                                            &                    \\
$M_{fin}$           &  The final mass of planets above which planetary growth is terminated (see Equation (\ref{M_max}))   &                    \\ 
$T_{m0}$          &   Midplane temperature at $r=R_*$ ($\simeq  ( 1/H )^{2/7} ( (\bar{C}_2 T_* R_*)/M_*)^{1/7} T_*$) &                \\ \hline
                          &  Model parameters$^1$                                                                                                               &                                    \\ \hline
$\rho_c$            &  The mean density of growing cores                                                                                           & 5.5 g cm$^{-3}$      \\
$b$                   &  A parameter for regulating the feeding zone                                                                               & 10                           \\
$m$                  &   The mass of planetesimals accreted by planetary cores                                                           & 10$^{18}$ g                 \\                  
$f_{fin}$            &  Final mass of planets (see Equation (\ref{M_max}))                                                                  & 10 ($>5$)                                  \\
$\Sigma_{A0}$  &  Surface density of active regions at $r=r_0$                                                                             & 20  g cm$^{-2}$ ($5 \leq \Sigma_{A0} \leq 50$) \\           
$s_A$                &  Power-law index of $\Sigma_A (\propto r^{s_A})$                                                                    & 3 ($1.5 \leq s_A \leq 6$)          \\
$\alpha_{A}$      &  Strength of turbulence in the active zone                                                                                 & $10^{-3}$ ($\alpha_A \leq 10^{-3}$)         \\        
$\alpha_{D}$     &  Strength of turbulence in the dead zone                                                                                   & $10^{-4}$ ($\alpha_D \leq 10^{-4}$)          \\  
$t$                     &  Power-law index of the disk temperature ($T \propto r^{t}$)                                                     & -1/2                               \\  
\hline
\end{tabular}

$^1$ As discussed in Appendix \ref{app1}, the variation of most parameters does not affect our results. 

The most uncertain parameter in the model may be $f_{fin}$ (see Section \ref{disc5}, also see HP13).
\end{center}
\end{minipage}
\end{table*}

The formation of planetary cores is determined by the growth timescale that is derived from the oligarchic growth, 
which is given as \citep[also see Table \ref{tableA1} for the definition of quantities]{ki02}
\begin{eqnarray}
\label{tau_cacc}
 \tau_{c,acc} & \simeq & 1.2 \times 10^{5} \mbox{ yr} \left( \frac{\Sigma_d}{10 \mbox{ g cm}^{-2}} \right)^{-1}
                    \left( \frac{r}{r_0} \right)^{1/2} \left( \frac{M_c}{M_{\oplus}} \right)^{1/3}
                    \left( \frac{M_*}{M_{\odot}} \right)^{-1/6} \\ \nonumber
             & \times & \left[ \left( \frac{b}{10} \right)^{-1/5}
                        \left( \frac{\Sigma_g}{2.4 \times 10^3 \mbox{ g cm}^{-3}} \right)^{-1/5} 
                       \left( \frac{r}{r_0} \right)^{1/20} \left( \frac{m}{10^{18} \mbox{ g}} \right)^{1/15}
                     \right]^2.
\end{eqnarray}
Two parameters are present in this timescale: $b$ and $m$. 
Detailed N-body simulations show that $b \simeq 10$ \citep[e.g.,][]{ki02} whereas the timescale is obviously insensitive to the value of $m$.
Adopting the timescale, planetary cores grow with time as follows;
\begin{equation}
 \frac{d M_p}{dt} =\frac{M_p}{\tau_{c,acc}}.
 \label{growth1}
\end{equation}

The core formation proceeds until the computed core mass, $M_{c}$, exceeds the critical mass of planetary cores, $M_{c,crit}$. 
As discussed in Section \ref{core_mass}, $M_{c,crit}$ is a function of the accretion rate of planetesimals by the cores.  
In the limit of the modest to high velocity dispersion $\sigma$ of planetesimals that are accreted onto the cores, 
$\dot{M}_c$ can be given as \citep{s72,il04i}
\begin{equation}
 \label{dotM_c}
 \dot{M}_c \sim 2 \pi \left( \frac{R_c}{r} \right) \left( \frac{M_c}{M_*} \right) 
                   \left( \frac{r\Omega}{\sigma}\right)^{2} \Sigma_d r^2 \Omega.  
\end{equation}
Since planetesimals within the feeding zones can reach cores when $\sigma/\Omega \sim \triangle r_c (=b r_H)$, 
$\dot{M}_c$ can be re-written as
\begin{equation}
  \label{mc_dot}
   \dot{M}_c  \sim  3.0\times 10^{-8} M_{\oplus} \mbox{ yr}^{-1} \left( \frac{b}{10} \right)^{-2} 
                 \left( \frac{\rho_c}{5.5 \mbox{ g cm}^{-3}} \right)^{-1/3} 
             \left( \frac{M_c}{M_{\oplus}} \right)^{2/3} \left( \frac{M_*}{M_{\odot}} \right)^{-1/3} 
                   \left( \frac{\Sigma_d}{10 \mbox{ g cm}^{-2}} \right)
                 \left( \frac{r}{r_0} \right) \left( \frac{\mbox{ yr}}{1 /\Omega} \right),
\end{equation}
where it is assumed that $R_c$ be similar to that of the Earth;  
\begin{equation}
 \label{r_c}
 R_c = 6.4 \times 10^{8} \mbox{ cm} \left( \frac{M_c}{M_{\oplus}}\right)^{1/3} 
                                   \left( \frac{\rho_c}{5.5 \mbox{ g cm}^{-3}} \right)^{-1/3}.
\end{equation}
In addition to $b$, Equation (\ref{mc_dot}) contains another parameter, $\rho_c$. 
Nonetheless, it is obvious that the choice of $\rho_c$ does not change the resultant value of $\dot{M}_c$ very much.   
In our computations, the value of $\dot{M}_c$ varies with time both through the orbital change of migrating planets 
as well as through the effects of time evolution of disks. 
Also, $\dot{M}_c$ is set zero once all the planetesimals in their feeding zones are consumed, 
and hence planetary cores attain the so-called isolation mass, which is defined by \citep{ki02,il04i} 
\begin{equation}
 \label{mc_iso}
 M_{c,iso} =  2 \pi r \triangle r_c \Sigma_d 
            \simeq  0.16 M_{\oplus} \left( \frac{b}{10} \right)^{3/2} 
                      \left( \frac{\Sigma_d}{10 \mbox{ g cm}^{-2}} \right)^{3/2} \left( \frac{r}{r_0} \right)^{3}
                      \left( \frac{M_*}{M_{\odot}} \right)^{-1/2}. 
\end{equation}

The subsequent gas accretion is already discussed in Section \ref{core_mass}, which is regulated by the Kelvin-Helmholtz timescale (see Equation (\ref{tau_KH})). 
The mass growth of planets is terminated when the planets obtain the final mass that is defined by (HP13)
\begin{equation}
 \label{M_max}
 M_{fin} \equiv   f_{fin} M_{gap},
\end{equation}
where the gap opening mass ($M_{gap}$) is given as
\begin{equation} 
 \label{M_gap}
          M_{gap} =   \mbox{min}\left[ 3 \left( \frac{H}{r} \right)^3, \sqrt{40 \alpha  \left( \frac{H}{r} \right)^5} \right] M_*.                              
\end{equation}
The choice of a free parameter, $f_{fin}(>1)$, is motivated by the recent numerical studies 
which show that a considerable amount of gas flows into the gap even after a clear gap is open in the gas disks \citep[e.g.,][]{lsa99,ld06}, 
which may lead to further growth of planets.  
A parameter study undertaken by HP13 shows that the resultant PFFs for the Jovian planets are insensitive to the value of $f_{fin}$ 
if $f_{fin} >5$ whereas those for the low-mass planets are very likely to be affected by varying $f_{fin}$ (also see Section \ref{disc5}).

\newlength{\myheight}
\setlength{\myheight}{0.5cm}
\newlength{\myheighta}
\setlength{\myheighta}{1cm}
\begin{table*}
\begin{minipage}{17cm}
\caption{Properties of planet traps}
\label{tableA2}
\begin{tabular}{cccc}
\hline
\parbox[c][\myheight][c]{0cm}{}  Planet trap           &  Position         &  Condition      &  Dust density   \\ 
\hline
\parbox[c][\myheighta][c]{0cm}{}  Dead Zone          & $\frac{r_{dz}}{r_0} = \left( \frac{\dot{M}}{3 \pi (\alpha_A+\alpha_D) \Sigma_{A0} H_0^2 \Omega_0} \right)^{1/(s_A+t+3/2)} $   &  N/A     &  $ \Sigma_{d,dz} \approx \frac{2 \dot{M} f_{dtg}}{3 \pi (\alpha_A+\alpha_D) H^2 \Omega}  $  \\   
\parbox[c][\myheighta][c]{0cm}{} Ice line                 &  $\frac{r_{il,\mbox{H}_2\mbox{O}}}{r_0} = \left[ \frac{1}{T_{m}^{12}(r_{il,\mbox{H}_2\mbox{O}})} \frac{\bar{C}_1 \bar{\kappa}_{il} \Omega_0^3} {\alpha_D}\left( \frac{\dot{M}}{3 \pi} \right)^2 \right]^{2/9} $  &   $ \frac{r_{il,\mbox{H}_2\mbox{O}}}{r_{dz}} >  \left( \frac{H}{r}(r_{dz}) \frac{\alpha_A + \alpha_D}{\alpha_A - \alpha_D} \right)^{1/(s_A+t/2+1)}$  & $ \Sigma_{d,il} \approx \frac{\dot{M} f_{dtg}}{3 \pi \alpha_A H^2 \Omega} $ \\
\parbox[c][\myheighta][c]{0cm}{} Heat transition     &  $ \frac{r_{ht}}{r_0}  = \left[ \frac{1}{T_{m0}} \left( \frac{r_0}{R_*} \right)^{3/7} \left( \frac{\bar{C}_1 \bar{\kappa}_{ht}\Omega_0^3}{\alpha_A} \left( \frac{\dot{M}}{3 \pi} \right)^2  \right)^{1/3}  \right]^{14/15} $ & $r_{ht}>r_{dz}$ &
$ \Sigma_{d,ht} \approx \frac{\dot{M} f_{dtg}}{3 \pi \alpha_A H^2 \Omega}$ \\
\hline
\end{tabular}

Note that there are five constants here: two of which are $\bar{C}_1 = 1.48 \times 10^{-4}$ in cgs units and $\bar{C}_2=5.38 \times 10^{14}$ in cgs units, 
another two of which are involved with the opacity law \citep{bl94}; $\bar{\kappa}_{il}= 2\times 10^{16}$ cm$^2$ g$^{-1}$ and 
$\bar{\kappa}_{ht} =2 \times 10^{-4}$ cm$^2$ g$^{-1}$, 
the last of which is the condensation temperature of water, $T_{m}(r_{il,\mbox{H}_2\mbox{O}})= 170$ K.      
Also see Tables \ref{table2} and \ref{tableA1} for the definition of variables.

\end{minipage}
\end{table*}

The orbital evolution of growing planets is prescribed by two kinds of planetary migration (HP12).
When protoplanets undergo core formation as well as slow gas accretion, 
the movement of planet traps determines the change of the position of the protoplanets. 
Specifically, the position of planet traps is used as that of trapped protoplanets (see Table \ref{tableA2}).
This occurs because the mass of protoplanets is generally smaller than the gap-opening mass (see Equation (\ref{M_gap})), 
so that the protoplanets are considered as rapid type I migrators that will be halted at planet traps (HP11).
It is assumed that planet traps are effective all the time for protoplanets whose masses are smaller than the gap-opening mass.

The properties of planet traps are summarized in Table \ref{tableA2} (also see Table \ref{tableA1}). 
Note that the dust density given there is used for calculating $\tau_{c,acc}$ and $\dot{M}_c$ when protoplanets are trapped at one of the traps.
Five parameters are required to define them: $\Sigma_{A0}$, $s_A$, $\alpha_A$, $\alpha_D$, $t$. 
Most of the parameters are relevant to the structure of dead zones (HP11).
Our previous study shows that the resultant PFFs are qualitatively similar when
$5 \leq \Sigma_{A0} \leq 50$, $1.5 \leq s_A \leq 6$, $\alpha_A \leq 10^{-3}$, and $\alpha_D \leq 10^{-4}$ (HP13).
For the value of $t$, many observations and theoretical calculations suggest that $t=-1/2$ \citep[e.g.,][]{hcg98,dccl98}.
Thus, a specific choice of these five parameters is very unlikely to affect our conclusions.

Once the protoplanets achieve the gap-opening mass, they drop-out from their host traps and undergo type II migration (HP12). 
There are two modes in type II migration, 
depending on the mass ratio of planets ($M_p$) to the total disk mass within the position of the planets ($2 \pi \Sigma_g r^2$) \citep[e.g.,][]{hi13}.
When $M_p < 2 \pi \Sigma_g r^2$, type II migration proceeds as the gas disks evolve.
As a result, the planets move inwards with the velocity 
written as
\begin{equation}
 v_{mig,II} \simeq - \frac{\nu}{r}.
\end{equation}
When the opposite situation is established, which generally occurs in the late stage of disk evolution, 
the type II migration rate slows down due to the inertia of the planets \citep{sc95,ipp99}. 
Eventually, the velocity of the planets becomes \citep{hi13}
\begin{equation}
 v_{mig,slowII} \simeq - \frac{\nu}{r} \frac{2 \pi \Sigma_g r^2}{M_p}.
 \label{slower_typeII}
\end{equation}
Note that the value of $\alpha_A$ ($\alpha_D$) is adopted for quantifying $\nu$ when the planets are beyond (within) dead zones.

\bibliographystyle{apj}          

\bibliography{apj-jour,adsbibliography}    

\end{document}